\documentclass[iop,apj,twocolumn]{emulateapj}
\usepackage{times}
\usepackage{epsfig}
\usepackage{amsmath, amsthm, amssymb}
\usepackage[plainpages=false, colorlinks=true, anchorcolor=blue,
  linkcolor=blue, citecolor=blue, bookmarks=false]
{hyperref}
\usepackage{color}
\usepackage{microtype}
\usepackage{float}
\usepackage[caption = false]{subfig}
\usepackage{graphicx}
\usepackage[export]{adjustbox}
\usepackage{diagbox}

\newcolumntype{$}{>{\global\let\currentrowstyle\relax}}
\newcolumntype{^}{>{\currentrowstyle}}

\graphicspath{{figures/}}

\begin{document}

\title{Impact of Pulsar and Fallback Sources on Multifrequency Kilonova Models}
\author{Ryan T. Wollaeger$^{1}$, Chris L. Fryer$^{1,2,3}$, Christopher J. Fontes$^{1}$,
  Jonas Lippuner$^{1}$, W. Thomas Vestrand$^{1}$, Matthew R. Mumpower$^{1,2,4}$,
  Oleg Korobkin$^{1,2}$, Aimee L. Hungerford$^{1}$, Wesley P. Even$^{1}$}

\affil{$^{1}$Center for Theoretical Astrophysics, Los Alamos National Laboratory,
  P.O. Box 1663, Los Alamos, NM 87545 \\
  $^{2}$Joint Institute for Nuclear Astrophysics - Center for the Evolution of the Elements, USA \\
  $^{3}$Department of Physics, The George Washington University, Washington, DC 20052 \\
  $^{4}$Theoretical Division, Los Alamos National Laboratory, Los Alamos, NM 87545}

\begin{abstract}

  We explore the impact of pulsar electromagnetic dipole and fallback accretion emission
  on the luminosity of a suite of kilonova models.
  The pulsar models are varied over pulsar magnetic field strength, pulsar lifetime,
  ejecta mass, and elemental abundances; the fallback models are varied over fallback
  accretion rate and ejecta mass.
  For the abundances, we use Fe and Nd as representatives of the wind and dynamical ejecta,
  respectively.
  We simulate radiative transfer in the ejecta in either 1D spherical or 2D cylindrical
  spatial geometry.
  For the grid of 1D simulations, the mass fraction of Nd is 0, $10^{-4}$, or $10^{-3}$
  and the rest is Fe.
  Our models that fit the bolometric luminosity of AT 2017gfo (the kilonova associated
  with the first neutron star merger discovered in gravitational waves, GW170817) do not
  simultaneously fit the B, V, and I time evolution.
  However, we find that the trends of the evolution in B and V magnitudes are better
  matched   by the fallback model relative to the pulsar model, implying the time
  dependence of the   remnant source influences the color evolution.
  Further exploration of the parameter space and model deficiencies is needed before
  we can describe AT 2017gfo with a remnant source.

\end{abstract}


\section{Introduction}
\label{sec:intro}

Under the standard accretion disk paradigm for gamma-ray bursts, the outflow is powered
by the release of accretion energy, typically assumed to be driven by magnetic fields.
Bursts were distinguished by their duration and hardness~\citep{kouveliotou1993}.
The burst duration in the accretion disk paradigm corresponds to the disk accretion timescale
which, in turn, corresponds to different progenitors~\citep{popham99}: long bursts are believed
to be produced in systems where the disk can be continuously fed (e.g. collapse of massive
stars) whereas short bursts are believed to be produced by compact disks (e.g. mergers of
compact binaries).
With these predictions for different progenitors for different bursts, theorists were able
to argue for different properties of short and long bursts with respect to their locations
in their host galaxies~\citep{bloom99,fryer99}.
Observations confirmed the distribution of locations~\citep{fong13}, verifying both the
compact binary progenitor and the accretion disk paradigm.
Neutron star/neutron star (NS/NS) and neutron star/black hole (NS/BH) mergers are the most
likely compact mergers behind these short bursts.

These compact mergers have also been invoked as the source of r-process elements, the
dynamically ejected material is so neutron rich that it produces a robust heavy
r-process yield.
Initially proposed over 4 decades ago~\citep{lattimer74}, increasingly detailed studies
support this as a leading source of r-process elements
\citep[e.g.][]{freiburghaus99,korobkin12,bauswein13a,lippuner15,radice16,thielemann17}.
For a review on the role of mergers in r-process production, see~\citep{cote2018}.
The merger also forms a disk of high angular momentum material that drives outflows
while accreting onto the central compact object.
This late-time outflow ejecta is bathed in neutrinos and is likely to be less neutron rich,
producing lighter (first peak r-process, iron peak) elements~\citep{metzger14}.
With the broadband and multi-messenger detection of GW170817
\citep{abbott2017a,abbott2017b,abbott2017c},
\footnote{See~\url{http://wise-obs.tau.ac.il/~arcavi/kilonovae.html} for the
  list of GW170817 discovery papers.}
astronomers were able to make the first definitive detection of a neutron star
merger with gravitational wave (GW) measurements providing proof of (and constraints on)
the merger, gamma-ray, X-ray and radio measurements of what appears to be a gamma-ray
burst jet and UVOIR measurements of the merger ejecta.

With these UVOIR measurements, astronomers can, for the first time, place observed
constraints on the r-process production in a neutron star merger.
Moreover, these observations place constraints on nuclear composition, and hence
nuclear mass models beyond the standard abundance curve~\citep{mumpower2016}.
The standard models for the UVOIR emission assume a broad range of ejecta including
both neutron-rich dynamical ejecta (producing heavy r-process, including lanthanides
that have high optical opacities) as well as lighter elements produced by the higher
electron-fraction ejecta during the accretion of the disk.
The radioactive decay of these elements powers a supernova-like light curve (kilonova) and,
by comparing this emission to the observations, astronomers can estimate the mass of the
ejecta.
The analysis of the emission to determine the exact ejecta from this merger depends
upon the opacities, the opacity implementation and the distribution of this ejecta, i.e.
composition, density, and velocity as a function of position (radius and angle).
Models ranged from constant opacity implementations to full lanthanide opacities
(utilizing a few representative isotopes) and distributions ranging from spherical mixes
to multi-component ejecta models.
None of the models capture all the physics and they predict a range of ejecta masses that
from the observations that varies by an order of magnitude~\citep{cote2018}.

If these uncertainties were not enough, additional power sources could
be augmenting the energy released from nuclear decay.
As the accretion rate lowers, the radiation from the material accreting on
the merged compact object is no longer trapped in the inflow and it
can augment the power of the burst.
In addition, there is some evidence that the merged core could be a
magnetized neutron star~\citep{li2018,piro2018}.
A normal, few times $10^{11}-10^{12} {\rm G}$ pulsar could power
the light curve.
Higher remnant magnetic fields, $\sim10^{14}-10^{16}$ G, have also been explored,
and can produce luminosity well into supernova range for $10^{-4}-10^{-2}$ M$_{\odot}$
\citep{yu2013,metzger2014}.
Alternatively, fallback~\citep{li2018,matsumoto2018} and cocoon
emission~\citep{kasliwal2017,matsumoto2018} have been invoked as supplying
additional power to the light curve of GW170817, on top of heating from r-process decay.
Similarly for GRB 130603B, motivated by the observed X-ray excess, a central X-ray
emission source undergoing reprocessing to the infrared has been used to lower ejecta
mass estimates from the IR excess~\citep{kisaka2016}.

In this paper, we present a grid of models to study the features of pulsar and
fallback accretion energy sources.
The outline of our models are given in Section~\ref{sec:models}.
Section~\ref{sec:ejecta} describes the 1- and 2-dimensional models used in
our calculations, Section~\ref{sec:remn} describes our remnant sources (pulsar
and accretion) and Section~\ref{sec:methods} describes the simulation methods
used to produce kilonova light curves.
With our grid of models, we produce a broad range of light curves and
spectra (Section ~\ref{sec:numres}).
Although we compare these to observations of GW170817, the intent
of this project is to provide a database of spectra and light curves
for upcoming observations.

\section{Model Properties and Simulations}
\label{sec:models}

\subsection{Ejecta Profiles}
\label{sec:ejecta}

For this project, we use both 1- and 2-dimensional outflow models.
To initialize the 1D spherical ejecta, we use a slight modification
to the semi-analytic, homologous solution discussed in Section 2.1.1
of~\cite{wollaeger2018},
\begin{subequations}
  \label{eq1:pmods}
  \begin{gather}
    v(r,t) = \frac{r}{t} \;\;, \\
    \rho(r,t) = \rho_0\left(\frac{t}{t_0}\right)^{-3}
    \left(1 - \frac{r^2}{(v_{\max}t)^2}\right)^3 \;\;, \\
    \mathcal{E}(r,t) = \mathcal{E}_0\left(\frac{t}{t_0}\right)^{-4}
    \left(1 - \frac{r^2}{(v_{\max}t)^2}\right)^4
    + E_{\rm source}(t)\frac{\delta(r)}{4\pi r^2} \;\;,
  \end{gather}
\end{subequations}
where $r$, $v$, $\rho$, $\mathcal{E}$, $t_0$, and $v_{\max}$ are
radius, velocity, density, radiation energy density, initial time, and
maximum velocity, respectively.
The value of $\mathcal{E}_0$ accounts for the contribution of r-process
nucleosynthesis and heating to the initial internal energy of the ejecta,
and is the same value used by~\cite{wollaeger2018}.
The value $E_{\rm source}$ is the energy contribution from a source near
the compact remnant (either from pulsar luminosity or accretion energy).
We will discuss this source in more detail in section~\ref{sec:remn}.
The composition of this material is assumed to be either dominated by
iron peak elements using Fe for the opacity  or mostly dominated by
iron peak elements with trace amounts of heavy r-process where lanthanide
opacities are the most critical (with a mass fraction of $10^{-4}$ or
$10^{-3}$ in Nd to represent lanthanides).
For the ejecta, we adopt masses and velocities similar to the
models of~\cite{li2018}.

\begin{figure}
  \centering
  \includegraphics[width=0.4\textwidth]{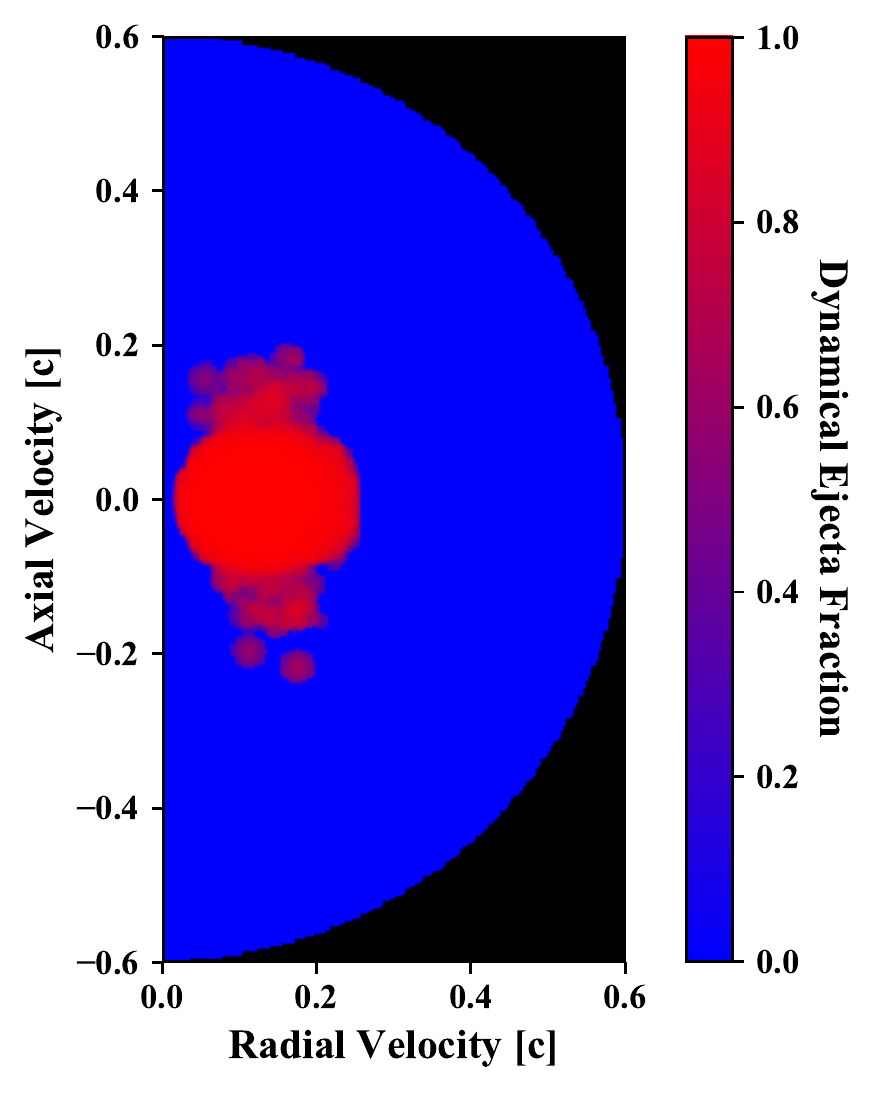}
  \caption{
    Fraction of dynamical ejecta at each point in the 2D morphology.
    The dynamical ejecta component (red) is from~\cite{rosswog2014}
    and the spherical wind component (blue) is from Eq.
    \eqref{eq1:pmods}~\citep{wollaeger2018}.
  }
  \label{fg1:ejecta}
\end{figure}

We also include a suite of models assuming a 2-component ejecta model
in 2D cylindrical geometry, superimposing the spherically symmetric wind
described above onto the ``model A'' dynamical ejecta of the SPH
simulations by~\cite{rosswog2014} (see also~\cite{rosswog2013}).
This is similar to the 2-component models of~\cite{wollaeger2018}
, though their wind velocity and mass are lower and higher,
respectively, following the simulations of~\cite{perego2014}.
The model A dynamical ejecta was derived from the simulation of the merger
of two 1.4 M$_{\odot}$ neutron stars, which produced an ejected mass of
0.013 M$_{\odot}$~\citep{rosswog2014}.
For the simulations in {\tt SuperNu}, this ejecta has been mapped to
an axisymmetric 2D grid; hence the 3D variations around the merger axis
are lost.
For the wind composition, we use the same abundance options as in our
1-dimensional models, Fe.
For the dynamical ejecta, we use Nd to represent a lanthanide-rich
ejecta.
The Model A ejecta proved to have an unobscured region permitting viewing
angles where a blue wind transient can manifest~\citep{wollaeger2018}.
Consequently, the viewing-angle dependence of the light curves and spectra
from this model permits the effect of the remnant luminosity to be observed
at different degrees of obscurity.
Figure~\ref{fg1:ejecta} has dynamical ejecta fraction at each velocity
coordinate, where it is colored red where dynamical ejecta dominates,
blue where wind ejecta dominates and black where there is no ejecta.
The fast wind can be seen to completely surround the dynamical ejecta, but
is of relatively low density.
Half of the wind mass is within a radius of $0.3c$, and the remnant source
is always at the origin.
We have not explored non-spherical wind morphologies, which may
affect the expression of the remnant source in the blue kilonova.

\subsection{Remnant Sources Enhancing the Kilonova Emission}
\label{sec:remn}

In most kilonova light-curve models, the emission is powered by the
decay of radioactive isotopes.
For our models, we include this energy, but we also include the energy
from an active remnant region:  either pulsar or accretion luminosity.
At early times, this energy is trapped in the outflowing ejecta and we
can treat it as an energy source, i.e. $E_{\rm source}$ in Eq.~(\ref{eq1:pmods}).
Consequently, we initialize our radiative transfer simulations by
solving for $E_{\rm source}$ at $t_0$, assuming the energy balance
is determined by adiabatic cooling and energy injection from the
remnant source.
After our initialization, the evolution of $\mathcal{E}(r,t)$ is
fully determined by radiative transfer.
Let's review the features of our pulsar and accretion luminosities,
and how the initial value of $E_{\rm source}$ is calculated.

\subsubsection{Pulsar Luminosity}
\label{sec:plum}

Especially in newly formed neutron stars like the merged compact
object in NS/NS binaries, pulsar emission can vary due to magnetic
field restructuring, high-order field configurations, etc.
Here we assume a simple dipole magnetic field source with a constant
magnetic field following the pulsar luminosity formulae explored in recent
studies~\citep{lasky2016,li2018,piro2018}.
Following~\cite{lasky2016}, the loss of neutron star remnant angular
kinetic energy is balanced by EM dipole and GW quadrupole luminosity,
\begin{equation}
  \label{eq1:plum}
  - I\Omega\dot{\Omega} = \frac{B_p^2R^6\Omega^4}{6c^3}
  + \frac{32GI^2\epsilon^2\Omega^6}{5c^2} \;\;,
\end{equation}
where the variables are defined by~\cite{lasky2016}.
The pulsar luminosity is given as the electromagnetic (EM) dipole luminosity multiplied by an
efficiency parameter, $\eta$, corresponding to the radiation beaming angle
\citep{rowlinson2014,lasky2016}.
Assuming GW dominated spin-down, the luminosity is~\citep{lasky2016}
\begin{equation}
  \label{eq4:plum}
  L_d(t) = \frac{\eta B_p^2R^6\Omega(t)^4}{6c^3}
  = L_0\left(1 + \frac{t}{t_{\rm gw}}\right)^{-1} \;\;,
\end{equation}
where
\begin{subequations}
  \label{eq5:plum}
  \begin{gather}
    L_0 = \frac{\eta\Omega_0^4B_p^2R^6}{6c^3} \;\;, \\
    t_{\rm gw} = \frac{5c^5}{128GI\epsilon^2\Omega_0^4} \;\;.
  \end{gather}
\end{subequations}

To explore the effect of variable remnant lifetime, we introduce a
time cut-off, $t_{\rm cut}$, to the luminosity,
\begin{equation}
  \label{eq6:plum}
  L_d(t) = L_0\left(1 + \frac{t}{t_{\rm gw}}\right)^{-1}\Theta(t_{\rm cut} - t)
  \;\;,
\end{equation}
where $\Theta$ is the unit step function.

The energy source $E_{\rm source}$ in Eq.~(\ref{eq1:pmods}) for our pulsar
models can then be written as:
\begin{equation}
  \label{eq2:pmods}
  \dot{E}_{\rm source} = -\frac{E_{\rm source}}{t} + L_d \;\;,
\end{equation}
where $L_d$ is given from Eq.~\eqref{eq6:plum} and $E_d(0) = 0$.
Note Eq.~\eqref{eq2:pmods} is a simplification of the ejecta layer
equations given by~\cite{metzger2017}, neglecting diffusion and
changes to the inner velocity from the impulse of the pulsar luminosity.
Thus we have assumed that the pulsar luminosity is not high enough to
significantly impact the morphology of the ejecta.
The solution to Eq.~\eqref{eq2:pmods} is
\begin{equation}
  \label{eq3:pmods}
  E_{\rm source}(t) = L_0\frac{t_{\rm gw}}{t}
  \left[t'-t_{\rm gw}\ln\left(\frac{t'+t_{\rm gw}}{t_{\rm gw}}\right)\right] \;\;,
\end{equation}
where $t' = \min(t,t_{\rm cut})$; the derivative with respect to $t'/t_{\rm gw}$
of the term in braces is always positive, showing $E_d(t) \geq 0$.

In our simulations, $E_d(t_0)$ is added to the innermost spatial cell
at a start time of $t_0$.  If $t_{\rm cut} > t_0$, the pulsar is still
active during the simulation, and Eq.~\eqref{eq6:plum} is integrated
over each time step to add further energy to the innermost cell.  In
all simulations, we use $t_0 = 10^4$ s.

\subsubsection{Fallback Luminosity}
\label{sec:flum}

Another source of energy can come from fallback after the initial kilonova explosion.
Fallback in stellar outbursts follows a simple power law with time
($\dot{m} \propto t^{-5/3}$ where $\dot m$ is the fallback accretion rate and $t$ is
the time)~\citep{chevalier89}.
The energy and mass ejected is more complex and high-resolution models have been
studied in the case of supernova fallback~\citep{fryer09}.
These models showed that roughly $\sim$10-25\% of the fallback matter is re-ejected,
carrying away roughly 10-25\% of the accretion energy.
These properties are directly applicable to the fallback in this scenario and
is similar to the simple prescriptions used in the kilonova community, e.g., \cite{li2018}:
\begin{equation}
  \label{eq1:flum}
  L_f(t) = \frac{\eta}{\eta_r} L_0\left(\frac{\dot{m}_0}{\dot{m}_r}\right)
  \left(\frac{t}{t_{\rm acc}}\right)^{-5/3}
  = \tilde{L}_0\left(\frac{t}{t_{\rm acc}}\right)^{-5/3} \;\;,
\end{equation}
where $\eta$ is again an efficiency parameter (roughly 10-25\%), $t_{\rm acc}$ is
initial time (when the luminosity begins to fall off), $\dot{m}_0$ is the initial
accretion rate, and $\dot{m_r}$ is a reference accretion rate.
Reference values $\eta_r$ and $\dot{m}_r$ are taken to be 0.1 and
$10^{-3}$ M$_{\odot}$/s, respectively.

As with our pulsar emission, we do not modify the ejecta mass or momentum in our
outflow, focusing instead on the energy injected from this accretion.
This luminosity corresponds to a source energy of
\begin{equation}
  \label{eq1:fmods}
  \dot{E}_{\rm source} = -\frac{E_{\rm source}}{t} + L_{\rm fallback} \;\;,
\end{equation}
and $L_{\rm fallback}$ is given by Eq.~\eqref{eq1:flum}.
Assuming the luminosity is proportional to $t^{-5/3}$ back to $t_{\rm acc}$, then
$E_{\rm source}$ can be solved analytically from Eq.~\eqref{eq1:fmods}.
We assume the luminosity $t^{-5/3}$ dependence holds as early as $t_{\rm acc}$
(which may only be true at $t\gg t_{\rm acc}$~\citep{metzger2017}).
Assuming the fallback luminosity is constant before $t_{\rm acc}$, the solution is
\begin{equation}
  \label{eq2:fmods}
  E_{\rm source} = \begin{cases}
    \displaystyle \tilde{L}_0\frac{t_{\rm acc}^2}{2t} + \tilde{L}_0t_{\rm acc}^{5/3}\left(\frac{1}{t^{2/3}}
    - \frac{t_{\rm acc}^{1/3}}{t}\right) \;\;,\;\; t>t_{\rm acc} \;\;, \\
    \displaystyle\frac{t_{\rm acc}\tilde{L}_0}{2} \;\;,\;\; t\leq t_{\rm acc} \;\;.
    \end{cases}
\end{equation}
Unlike the pulsar models, we do not introduce a cut-off time for the source.

For the fallback models, a considerable amount of energy can be injected
into the ejecta on short time scales.
Adopting parameters similar to~\cite{li2018}, $\eta=0.1$, $L_0=2\times10^{51}$ erg/s,
and $\dot{m}_0=10^{-3}$ M$_{\odot}$/s, from Eq.~\eqref{eq2:fmods}, the energy
added up to $t_{\rm acc}$ is $10^{50}$ erg.
The kinetic energy of our smallest-mass ejecta is
$\sim M_{\rm ej}(v_{\max}/2)^2/2\approx10^{50}$ erg.
Assuming all the energy before $t_{\rm acc}$ goes into boosting the kinetic energy,
the resulting average velocity should be increased to about
$\tilde{v}_{\max}/2=\sqrt{2}v_{\max}/2\approx0.42c$.
Consequently, we also simulate a suite of fallback kilonova models with
$v_{\max}/2=0.45$ and with
\begin{equation}
  \label{eq3:fmods}
  E_{\rm source} = \begin{cases}
    \displaystyle \tilde{L}_0t_{\rm acc}^{5/3}\left(\frac{1}{t^{2/3}}
    - \frac{t_{\rm acc}^{1/3}}{t}\right) \;\;,\;\; t>t_{\rm acc} \;\;, \\
    0 \;\;,\;\; t\leq t_{\rm acc} \;\;,
  \end{cases}
\end{equation}
as an alternative to adding the energy for $t<t_{\rm acc}$ as a radiative source.
We must note that this adjustment to the model does not fully account
for the morphological effects of the early source, which would squeeze the
ejecta to produce a morphology more similar to those of~\cite{metzger2017} and
\cite{li2018}.

\subsection{Methods}
\label{sec:methods}

For the radiative transfer, we use {\tt SuperNu}
\citep{wollaeger2013,wollaeger2014} with tabular opacities from
the LANL suite of atomic physics codes
\citep{fontes2015b,fontes2017}.
The opacity tables used here have been described by~\cite{fontes2019}.
For the bulk of our calculations, we use single elements to represent
the material:  Fe to represent the iron peak elements, Nd to represent
lanthanides.
For improved robustness in pre-peak luminosity kilonovae simulations
with high group resolution ($N_g\gtrsim1000$,
$\lambda\in[10^{3},1.28\times10^5]$ $\AA$), we have added a more rigorous
Doppler shift treatment for the diffusion optimization
\citep{densmore2012,abdikamalov2012,cleveland2014} in {\tt SuperNu}
(Wollaeger et al 2019, in prep).

The radiative transfer is semi-relativistic and, hence, only correct to
$O(v/c)$, which is a limitation that arises from the current implementation
of the diffusion optimization.
This issue is seemingly problematic for the pulsar kilonova models, which use
low mass, high-velocity (0.3-0.45$c$ median) ejecta.
However, a compensating phenomenon is the recession of the photosphere,
which tends to relegate radiation-matter interaction (where boosting is
important) to lower velocity values.
For our ejecta, we may estimate the photospherical recession with
\begin{equation}
  \label{methods:eq1}
  1 = \left(\frac{t_0^3\rho_0v_{\max}\kappa}{t^2}\right)\int_x^1(1-(x')^2)^3dx'
  \;\;,
\end{equation}
where $\kappa$ is a grey estimate of the opacity and $x = v/v_{\max}$.
The integral in Eq.~\eqref{methods:eq1} is analytic, and the resulting
expression can be solved for $x$ given $t$ (for instance, with Newton-Raphson
iteration) or for $t$.
For instance, for $\kappa=0.1$ and 1 cm$^2$/g, assuming $v_{\max} = 0.9c$, the
time at which the photosphere reaches $v = 0.1c$ is $t\approx 0.14$ or 0.45
days, respectively.
This would suggest that, on average, $O(v/c$) radiative transfer becomes
accurate on time scales relevant to observation of the wind.
Additionally, for the model pulsar in particular, the spin-down emission
should not be greatly impacted by the outflow speed, since the source is
located at the center of the ejecta.

The radiative transfer simulations employ 64 uniform spatial cells from
$v=0$ to $v=0.6c$, 400 logarithmic time steps from $10^4$ s to 20 days, and
1024 logarithmic wavelength groups from $10^3$ to $1.28\times10^5$ $\AA$.
The opacities are calculated on the same density-temperature grid as
of~\cite{wollaeger2018}: 17 logarithmic density points from $10^{-20}$ to
$10^{-4}$ g cm$^{-3}$, 27 temperature points from $0.01$ to $5$ eV, and 14,900
frequency points from $h\nu/kT=1.25\times10^{-3}$ to $h\nu/kT=3\times10^{4}$ for
each density and temperature.
The opacity frequency grid is mapped to the radiative transfer wavelength grid
by direct (unweighted) integral averaging.

Following the labeling conventions of~\cite{wollaeger2018}, we call models
SAFe (``semi-analytic (ejecta) with Fe'') and SAFeNd.
Otherwise, we exclude the other parameter variations (mass, pulsar magnetic
field, etc.) in the name, and write these out explicitly.

\section{Numerical Results}
\label{sec:numres}

With a range of ejecta properties (1- and 2-dimensional geometries, ejecta masses,
and different compositions), we can study the role of our two energy sources,
pulsars and fallback accretion.
In Sections~\ref{sec:rcut}-\ref{sec:pulsr2d}, we explore the effect of variations
of remnant lifetime, elemental abundances, ejecta mass, and magnetic field strength
on our pulsar kilonova models.
In Sections~\ref{sec:fvar}-\ref{sec:pulsr2d}, we vary the ejecta mass and accretion
rate in our fallback kilonva models.
Each variation has consequences for the observables from our models, which we
discuss in the sections that follow.

\subsection{Pulsar-Powered Light Curves}
\label{sec:pulsar_results}

With our pulsar model, we have a number of free parameters.
Here we study the kilonova light curves and spectra varying both the pulsar cutoff
timescale and the magnetic field strength.
We include both 1- and 2-dimensional geometries and vary the composition.

\subsubsection{Remnant Cutoff Times}
\label{sec:rcut}

We test the effect of varying the pulsar lifetime in 1D spherical models of wind-like
outflow, assuming Fe for the wind opacity.
The model parameters are summarized in Table~\ref{tb1:rcut}.

\begin{table}[H]
  \centering
  \caption{Parameters for test of remnant lifetime.}
  \label{tb1:rcut}
  \resizebox{0.75\columnwidth}{!}{
  \begin{tabular}{c|c}
    \hline
    $t_{\rm cut}$ & $\{0.1, 0.2, 1, 2, 4\}\times10^5$ s \\
    $M_{\rm ej}$ & $10^{-3}$ M$_{\odot}$ \\
    $v_{\max}/2$ & $0.3\,c$ \\
    $R$ & $1.2\times10^6$ cm \\
    $I$ & $1.5\times10^{45}$ g cm$^2$ \\
    $\Omega_0$ & $2\pi\times10^3$ s$^{-1}$ \\
    $B$ & $3.4\times10^{12}$ G \\
    $\epsilon$ & $0.0035$ \\
    $\eta$ & $1$ \\
    $(\chi_{\rm Fe}, \chi_{\rm Nd})$ & $(1, 0)$ \\
    \hline
  \end{tabular}
  }
\end{table}

The parameters for the pulsar imply $t_{\rm gw} = 495$ s and
$L_0 = 3.33\times10^{44}$ erg/s, similar to~\cite{li2018}.

\begin{figure}
    \centering
    \includegraphics[width=0.5\textwidth]{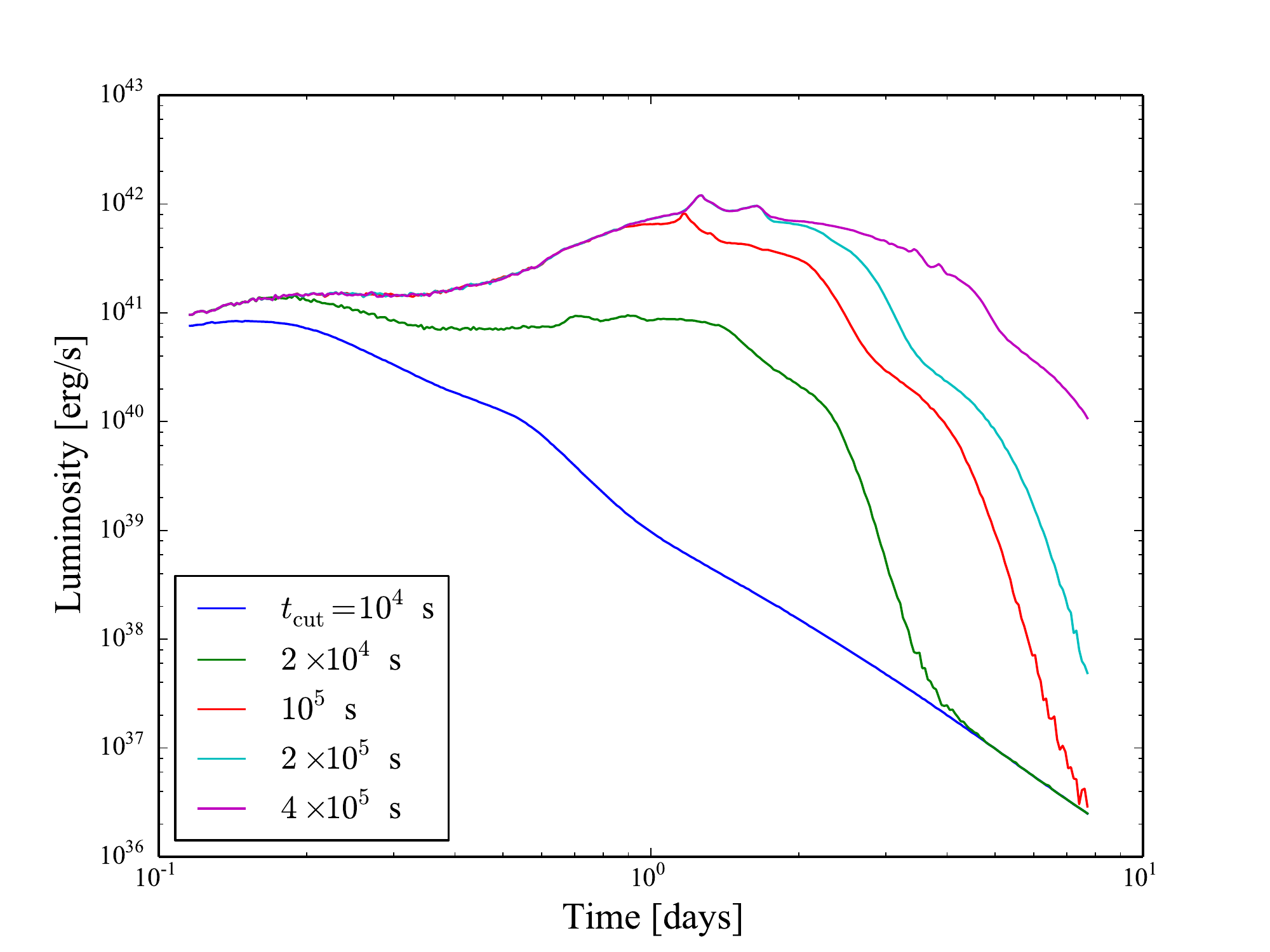}
    \caption{Bolometric luminosity versus time for the
      pulsar-KN models with variable remnant cutoff time.
      The parameters are given in Section~\ref{sec:rcut}
      Increasing the cutoff time, $t_{\rm cut}$, to values greater
      than 1 day does not greatly affect the peak luminosity.}
    \label{fg1:rcut}
\end{figure}

Figure~\ref{fg1:rcut} shows the bolometric luminosity versus time for these cutoff
time variations (see model parameters above).
The maximum peak luminosity is achieved for $t_{\rm cut}\gtrsim2\times10^5$ s;
increasing the remnant lifetime further only affects the brightness of the tail
of the light curve.
For $t_{\rm cut}\lesssim1\times10^5$ s ($\sim$day), the peak luminosity is
more sensitive to the remnant lifetime.
This sensitivity can be seen in the change of the 1-day luminosity with respect
to the $t_{\rm cut}$.

For sufficiently low cutoff times, the effect of the pulsar luminosity
becomes small relative to the r-process heating.
Notably, in Fig.~\ref{fg1:rcut} the light curve for the lowest cutoff time
appears to be monotonically decreasing.
This is an effect of the fast expansion speed, low mass, low opacity and
choice of morphology.
In particular, for this morphology the time of peak bolometric luminosity
follows~\citep{wollaeger2018}

\begin{multline}
  \label{eq1:rcut}
  t_{\rm peak} \approx \\ \left(1 \text{ day}\right)
  \left(\frac{\kappa}{10 \text{ cm}^2/\text{g}}\right)^{0.35}
  \left(\frac{M_{\rm ej}}{10^{-2} \text{ M}_{\odot}}\right)^{0.318}
  \left(\frac{v_{\max}}{0.2c}\right)^{-0.6} \;\;.
\end{multline}

Equation~\eqref{eq1:rcut} gives $t_{\rm peak}\approx0.05$ day, assuming $\kappa=0.1$
cm$^2$/g.
Alternatively, using the scaling relation for time of peak bolometric luminosity
from~\cite{grossman2014}, $t_{\rm peak}\approx0.09$ day.
The earlier peak time relative to the models of~\cite{grossman2014} is due to
the thermal energy contribution to the light curve~\citep{wollaeger2018}.

\subsubsection{Magnetic Field Strength Variations and Ejecta Mass and Composition}
\label{sec:vars}

With a better understanding of the role of the cutoff time, we can now study
the dependence of the pulsar-powered light curves on the magnetic field strength
for a range of ejecta masses and two different compositions using just two cutoff
times:  $2\times10^4$, $2\times10^5$ s.  We vary the composition by adding a small mass fraction of Nd.
Apart from the ejecta mass, magnetic field strength, and composition, the model parameters
are the same as in Section~\ref{sec:rcut}.
The model parameters are listed in Table~\ref{tb2:vars}.

\begin{table}[H]
  \centering
  \caption{Parameters for test of mass and magnetic field strength variation.}
  \label{tb2:vars}
  \resizebox{0.75\columnwidth}{!}{
  \begin{tabular}{c|c}
    \hline
    $t_{\rm cut}$ & $\{0.2, 2\}\times10^5$ s \\
    $M_{\rm ej}$ & $\{1, 3, 10\}\times10^{-3}$ M$_{\odot}$ \\
    $v_{\max}/2$ & $0.3\,c$ \\
    $R$ & $1.2\times10^6$ cm \\
    $I$ & $1.5\times10^{45}$ g cm$^2$ \\
    $\Omega_0$ & $2\pi\times10^3$ s$^{-1}$ \\
    $B$ & $\{1, 10, 100\}\times10^{12}$ G \\
    $\epsilon$ & $0.0035$ \\
    $\eta$ & $1$ \\
    $(\chi_{\rm Fe}, \chi_{\rm Nd})$ & $\{(1,0) \,,\, (1-10^{-4}, 10^{-4})\}$ \\
    \hline
  \end{tabular}
  }
\end{table}

\begin{figure*}
    \centering
    \subfloat[]{\includegraphics[width=0.5\textwidth]{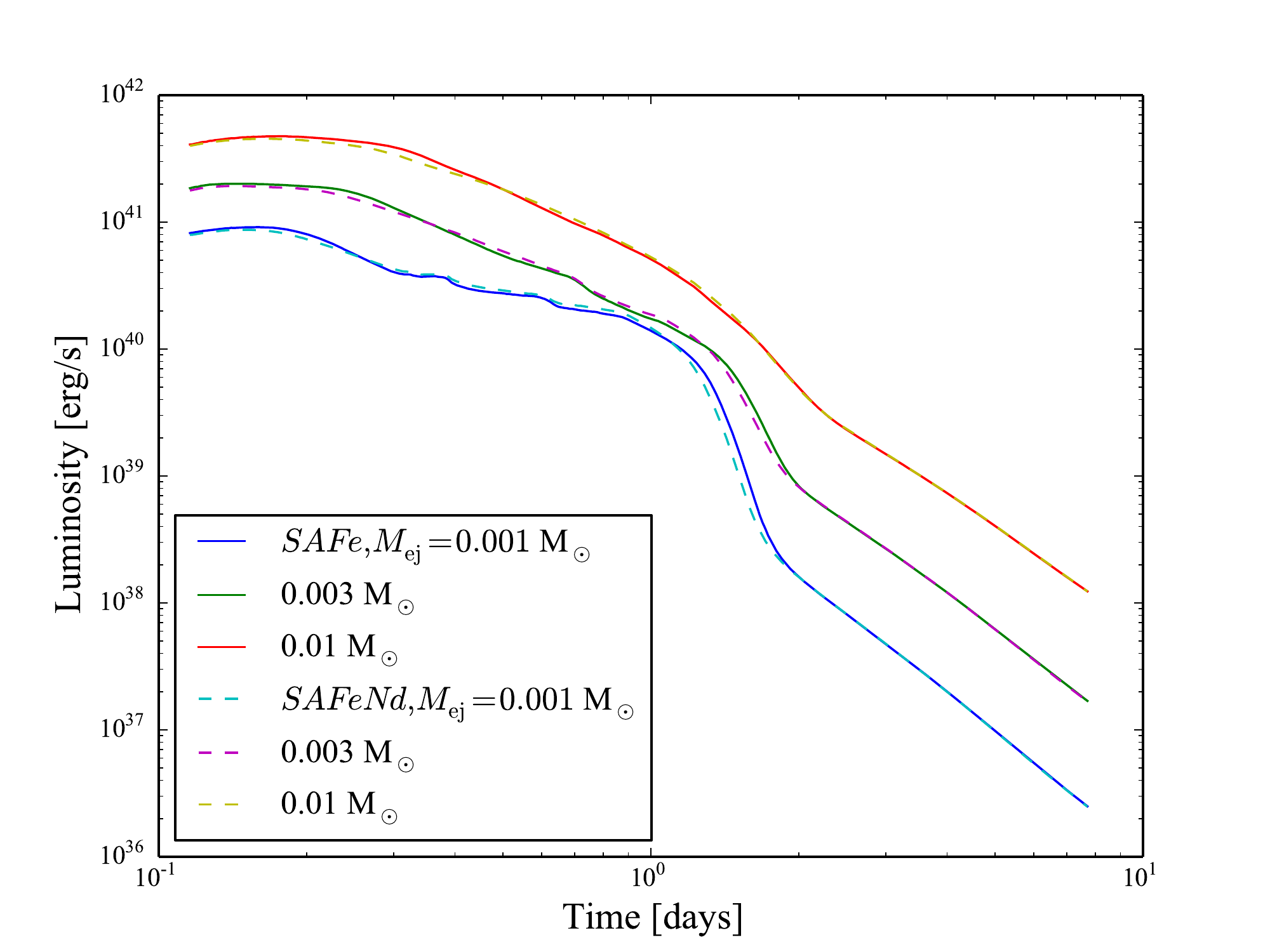}\label{fg1a:vars}}
    \subfloat[]{\includegraphics[width=0.5\textwidth]{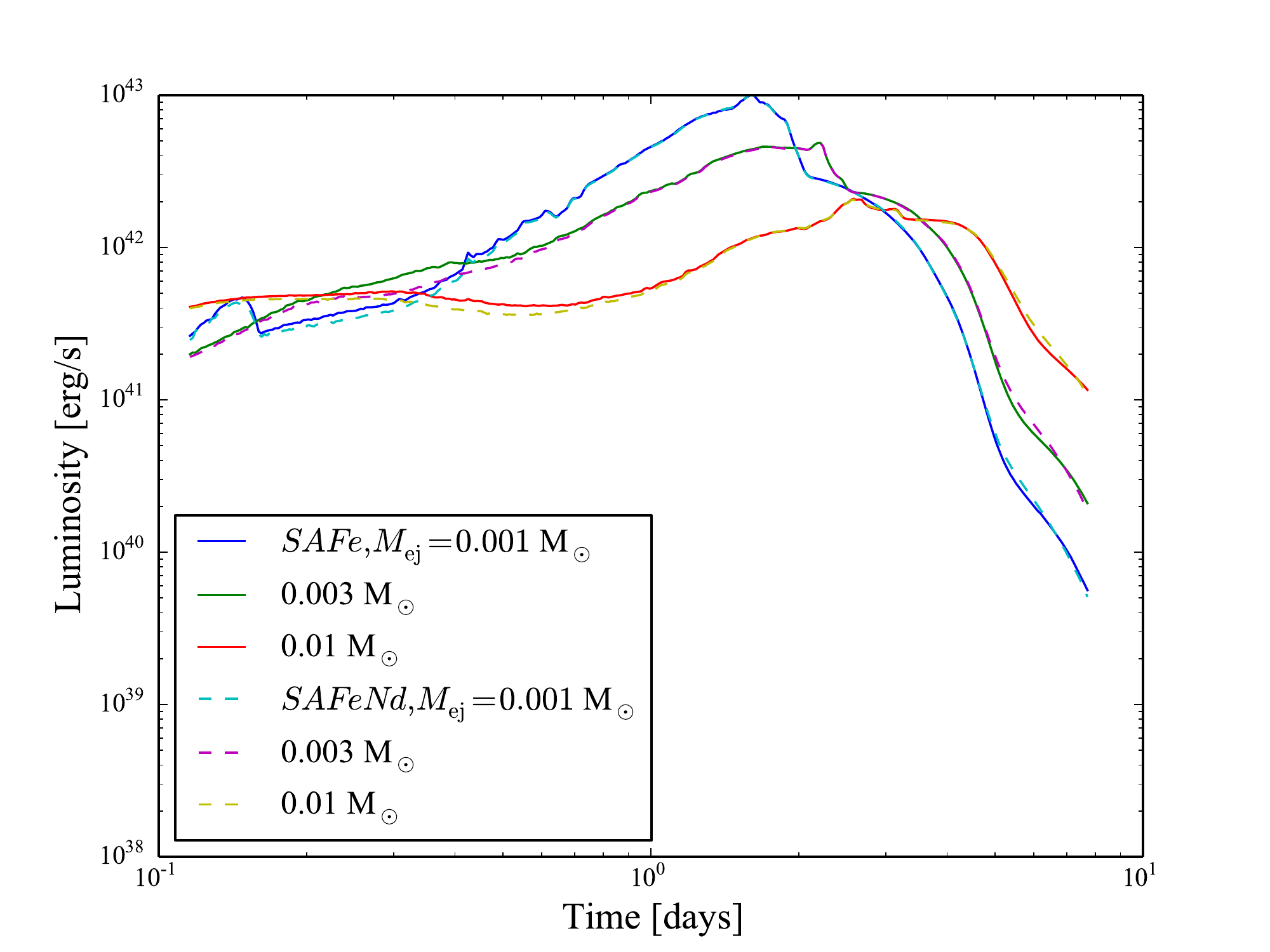}\label{fg2a:vars}} \\
    \subfloat[]{\includegraphics[width=0.5\textwidth]{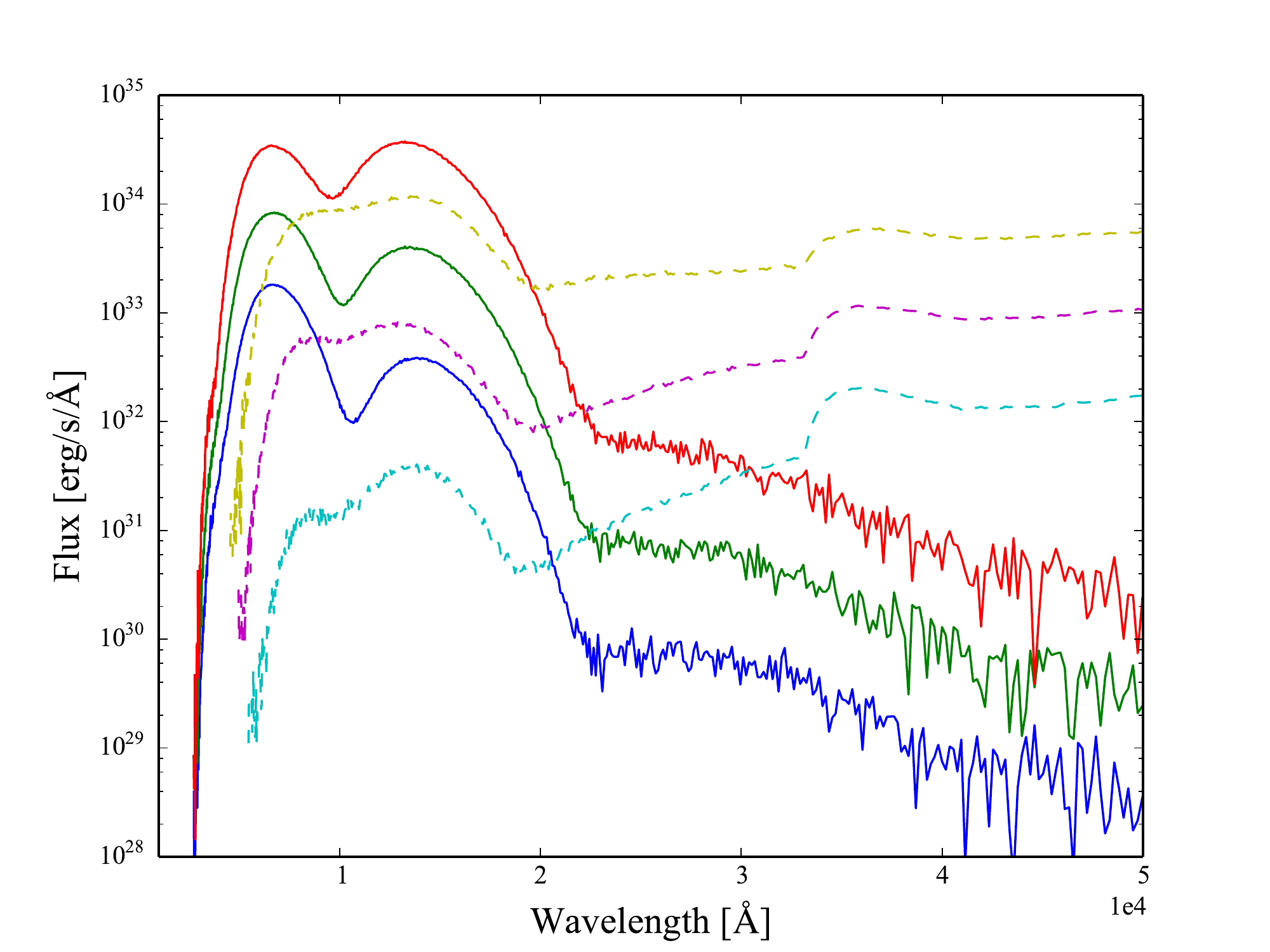}\label{fg1b:vars}}
    \subfloat[]{\includegraphics[width=0.5\textwidth]{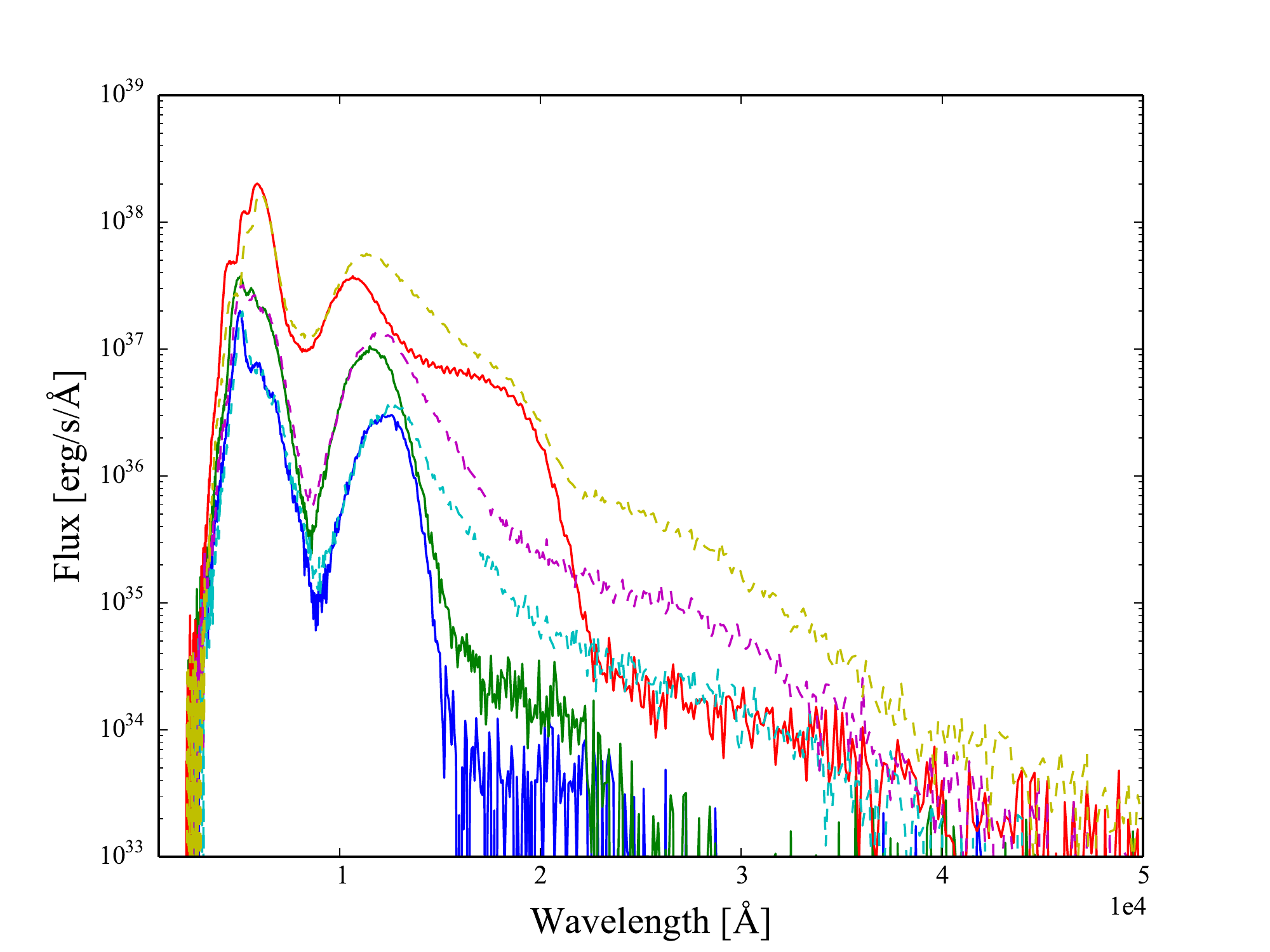}\label{fg2b:vars}}
    \caption{In Fig.~\ref{fg1a:vars}, bolometric luminosity versus time for each mass,
      for $B=10^{12}$ G and $t_{\rm cut}=2\times10^4$ s.
      In Fig.~\ref{fg1b:vars}, corresponding spectra at day 5.45.
      The models with mass fraction $\chi_{\rm Nd}=10^{-4}$ have redder emission, despite having similar
      bolometric luminosity; the blue features corresponding to Fe diminish more quickly after about 2 days.
      In Fig.~\ref{fg2a:vars}, bolometric luminosity versus time for each mass,
      for $B=10^{13}$ G and $t_{\rm cut}=2\times10^5$ s.
      In Fig.~\ref{fg2b:vars}, corresponding spectra at day 5.45.
      The persistent, brighter pulsar luminosity model seems to support the blue features of the
      spectrum persisting longer.}
    \label{fg1:vars}
\end{figure*}

The parameters for the pulsar again imply $t_{\rm gw} = 495$ s; the
pulsar luminosity is $L_0\in\{2.88, 288, 28800\}\times10^{43}$ erg/s.
Tables~\ref{tb3:vars} and~\ref{tb4:vars} display the luminosity at day 1 for
the SAFe models with $2\times10^4$ and $2\times10^5$ s, respectively.
Tables~\ref{tb5:vars} and~\ref{tb6:vars} show the same data for the
SAFeNd models.
For the lowest magnetic field strength, $10^{12}$ G, and lowest remnant
cutoff time, increasing the mass produces an upward trend in the luminosity
at day 1.
For these parameters, the r-process decay energy competes with the pulsar
luminosity in setting the kilonova bolometric luminosity at day 1.
This is discernible in Fig.~\ref{fg1a:vars}, where increasing the ejecta
mass can be seen to mask the pulsar peak around day 1.
With a remnant lifetime of $2\times10^5$ s, increasing the ejecta mass from
0.001 to 0.003 M$_{\odot}$ lowers the luminosity at day 1.
The diminished luminosity at day 1 indicates delay in the emission from
increased optical depth.
For the higher magnetic field strengths, the luminosity at day 1 only
decreases when more mass is added, resulting from the increased optical
depth.
The dimming and delaying of the peak luminosity for the models with
$B=10^{13}$ G and $t_{\rm cut}=2\times10^5$ s can be seen in
Fig.~\ref{fg1b:vars}.

The introduction of the Nd mass fraction, $\chi_{\rm Nd}=10^{-4}$, does not
appear to significantly impact the bolometric luminosity, with respect to
the pure Fe models.
At sufficiently late time, the bolometric luminosity is set by the r-process
decay rate.
However, the spectrum at later times tends to be redder; in particular for
the low-pulsar luminosity, low-remnant time cutoff spectrum shown in
Fig.~\ref{fg1b:vars}, the spectral features corresponding to the blue
transient are systematically dimmer with $\chi_{\rm Nd}=10^{-4}$.
In Fig.~\ref{fg2b:vars}, the brighter, longer duration pulsar luminosity
appears to sustain the blue portion of the spectrum further in time.

\begin{table}[H]
  \centering
  \caption{SAFe luminosity (erg/s) at day 1 with $t_{\rm cut}=2\times10^4$ s.}
  \label{tb3:vars}
  \resizebox{1.0\columnwidth}{!}{
  \begin{tabular}{|c|ccc|}
    \hline
    \backslashbox{$M_{\rm ej}$ (M$_{\odot}$)}{$B$ (G)} & $10^{12}$ & $10^{13}$ & $10^{14}$ \\
    \hline
    0.001 & $1.41\times10^{40}$ & $7.13\times10^{41}$ & $3.11\times10^{43}$ \\
    0.003 & $1.74\times10^{40}$ & $2.60\times10^{41}$ & $4.65\times10^{42}$ \\
    0.01 & $5.17\times10^{40}$ & $1.40\times10^{41}$ & $1.93\times10^{42}$ \\
    \hline
  \end{tabular}
  }
\end{table}

\begin{table}[H]
  \centering
  \caption{SAFe luminosity (erg/s) at day 1 with $t_{\rm cut}=2\times10^5$ s.}
  \label{tb4:vars}
  \resizebox{1.0\columnwidth}{!}{
  \begin{tabular}{|c|ccc|}
    \hline
    \backslashbox{$M_{\rm ej}$ (M$_{\odot}$)}{$B$ (G)} & $10^{12}$ & $10^{13}$ & $10^{14}$ \\
    \hline
    0.001 & $9.96\times10^{40}$ & $4.55\times10^{42}$ & $8.52\times10^{43}$ \\
    0.003 & $6.48\times10^{40}$ & $2.34\times10^{42}$ & $3.67\times10^{43}$ \\
    0.01 & $8.48\times10^{40}$ & $5.36\times10^{41}$ & $1.91\times10^{43}$ \\
    \hline
  \end{tabular}
  }
\end{table}

\begin{table}[H]
  \centering
  \caption{SAFeNd luminosity (erg/s) at day 1 with $t_{\rm cut}=2\times10^4$ s.}
  \label{tb5:vars}
  \resizebox{1.0\columnwidth}{!}{
  \begin{tabular}{|c|ccc|}
    \hline
    \backslashbox{$M_{\rm ej}$ (M$_{\odot}$)}{$B$ (G)} & $10^{12}$ & $10^{13}$ & $10^{14}$ \\
    \hline
    0.001 & $1.49\times10^{40}$ & $7.02\times10^{41}$ & $3.10\times10^{43}$ \\
    0.003 & $1.88\times10^{40}$ & $2.65\times10^{41}$ & $4.52\times10^{42}$ \\
    0.01 & $5.37\times10^{40}$ & $1.39\times10^{41}$ & $1.76\times10^{42}$ \\
    \hline
  \end{tabular}
  }
\end{table}

\begin{table}[H]
  \centering
  \caption{SAFeNd luminosity (erg/s) at day 1 with $t_{\rm cut}=2\times10^5$ s.}
  \label{tb6:vars}
  \resizebox{1.0\columnwidth}{!}{
  \begin{tabular}{|c|ccc|}
    \hline
    \backslashbox{$M_{\rm ej}$ (M$_{\odot}$)}{$B$ (G)} & $10^{12}$ & $10^{13}$ & $10^{14}$ \\
    \hline
    0.001 & $1.00\times10^{41}$ & $4.52\times10^{42}$ & $8.38\times10^{43}$ \\
    0.003 & $6.86\times10^{40}$ & $2.31\times10^{42}$ & $3.63\times10^{43}$ \\
    0.01 & $8.57\times10^{40}$ & $5.10\times10^{41}$ & $1.81\times10^{43}$ \\
    \hline
  \end{tabular}
  }
\end{table}

We have tabulated the day 1 and day 7 bolometric luminosities and UBVRIJHK broadband
magnitudes for these models in Tables~\ref{tb1:tables}-\ref{tb4:tables} of the Appendix.
The intent of tabulating the data at early and later time is to identify the effect of
variations in the model on evolution of the luminosity and magnitudes.
As an example, we can see that the sensitivity of the day 7 model data to remnant cutoff
time depends on the pulsar magnetic strength.
For instance, in Table~\ref{tb3:tables}, for $B=10^{12}$ G and $M_{\rm ej}=0.001$ M$_{\odot}$,
the B-band magnitude is 32.9 at day 7 for models with either $t_{\rm cut}=2\times10^4$ or
$2\times10^5$ s.
The same models with $B=10^{13}$ G have day 7 B-band magnitudes of 32.9 and 25.3 for
$t_{\rm cut}=2\times10^4$ and $2\times10^5$ s, respectively.

\subsubsection{2-component, 2-dimensional Pulsar Results}
\label{sec:pulsr2d}

\begin{figure}
  \centering
  \includegraphics[width=0.45\textwidth]{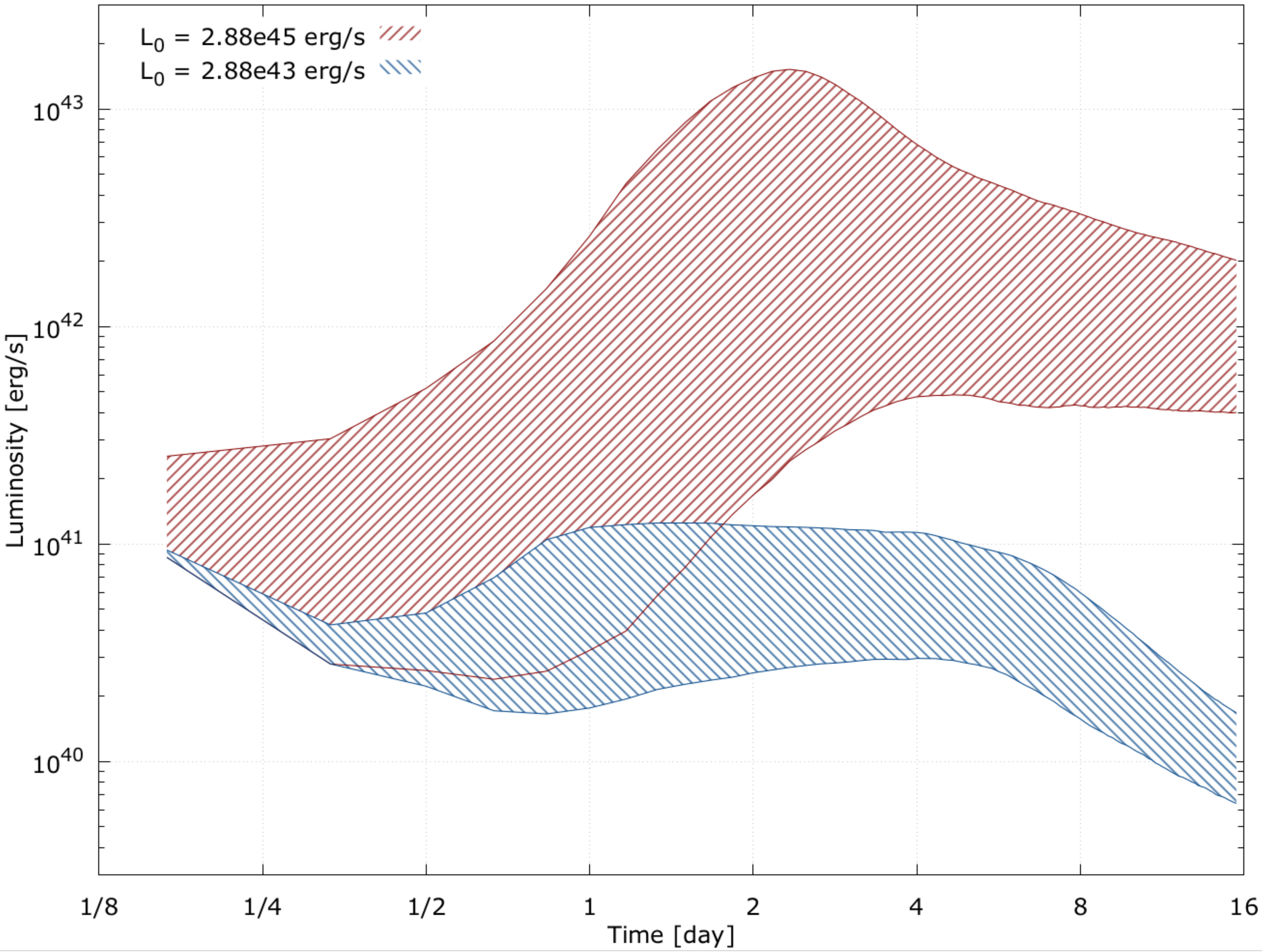}
  \caption{Isotropically equivalent bolometric luminosity versus
    time for the 2D pulsar kilonova models
    described in Fig.~\ref{fg1:ejecta} and Section~\ref{sec:pulsr2d}.
    The shaded regions represent the range of luminosities from the 54 angular
    bins that are uniform in the cosine of the polar viewing angle.
    Brighter luminosity corresponds to viewing bins that are more aligned with
    the merger axis.
  }
  \label{fg1:pulsr2d}
\end{figure}

The morphology of the dynamical ejecta of the 2D models is ``model A''
used by~\cite{wollaeger2018}.
The wind superimposed on the model A ejecta has the properties described
in Section~\ref{sec:rcut} for $B=10^{12}$ and $10^{13}$ G, but without a remnant cutoff time.
These field strengths correspond to initial pulsar luminosities of $2.88\times10^{43}$
and $2.88\times10^{45}$ erg/s, respectively.
Figure~\ref{fg1:pulsr2d} has bolometric luminosity for the range of angular views as shaded
regions for each model.
These luminosities are ``isotropic equivalents'', where each is divided by the solid
angle per view and multiplied by 4$\pi$.
The angular views are divided into 54 polar angular viewing ranges of equal solid angle.
The dimmest light curves (or lowest edge of the shaded regions) correspond to views most
closely aligned with the merger plane (``edge-on''), which are most obscured by Nd
in the dynamical ejecta.
The axial view is over an order of magnitude brighter than for the edge-on view for the
model with $B=10^{13}$ G.

\begin{figure*}
    \centering
    \subfloat[]{\includegraphics[width=0.5\textwidth]{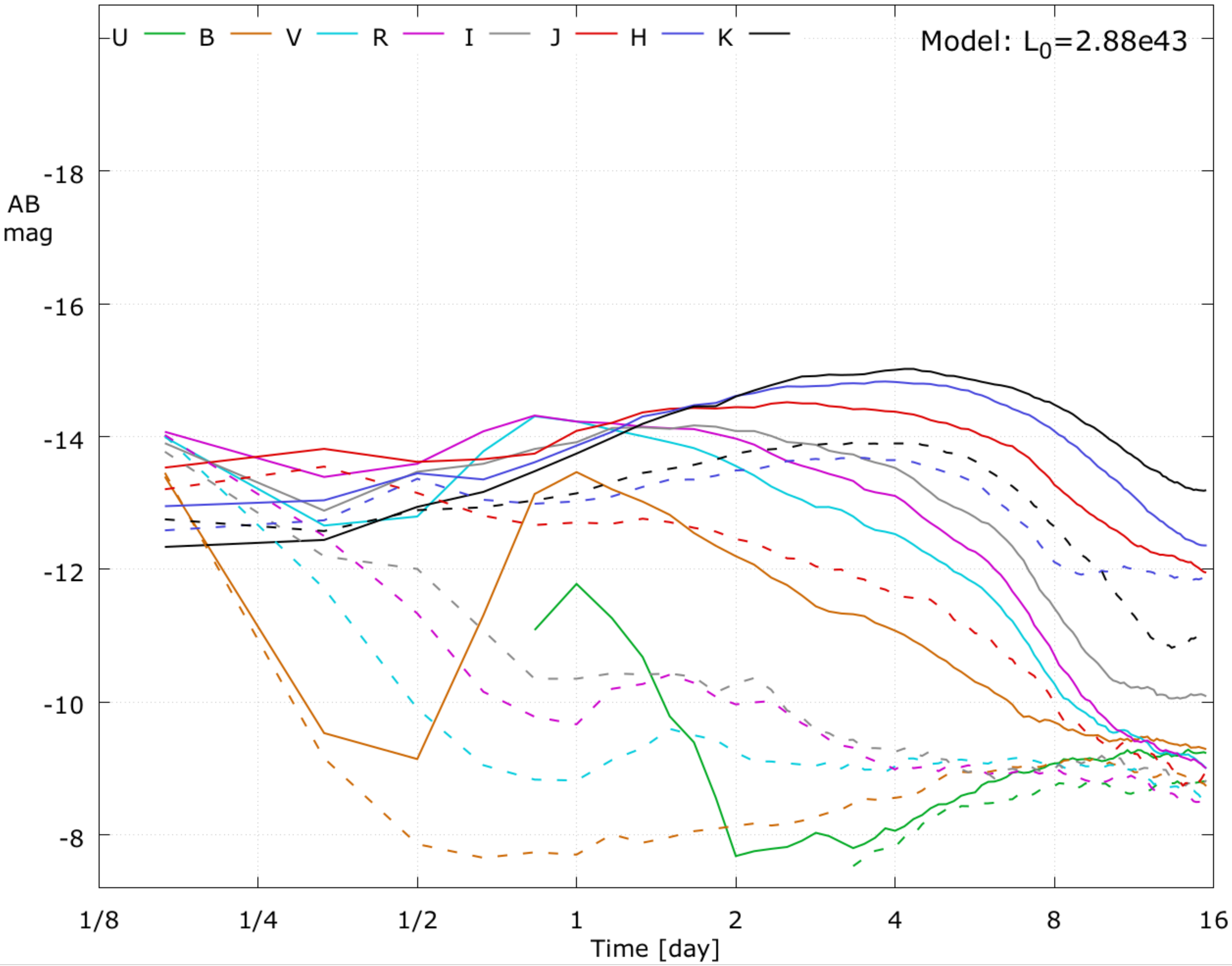}\label{fg2a:pulsr2d}}
    \subfloat[]{\includegraphics[width=0.5\textwidth]{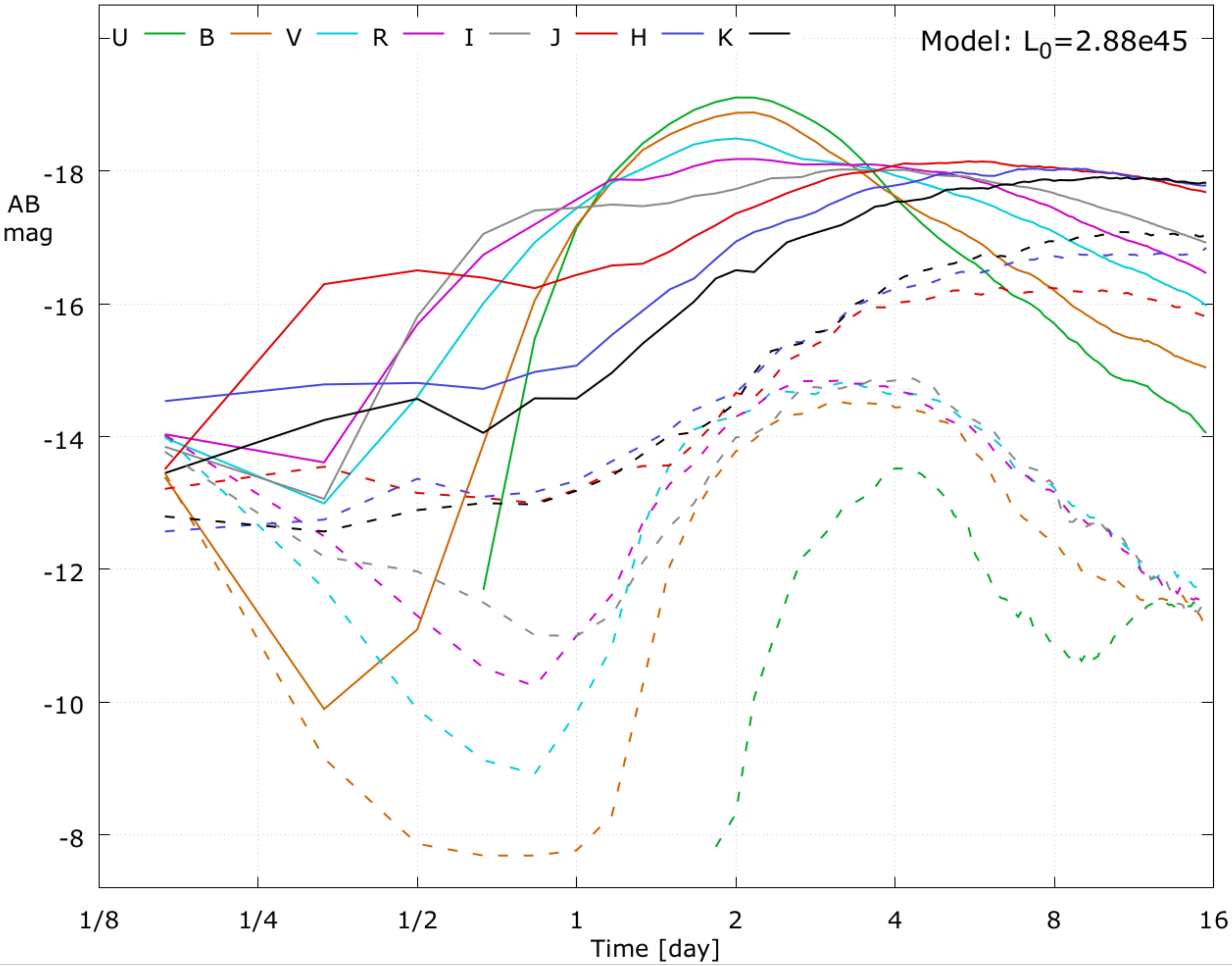}\label{fg2b:pulsr2d}}
    \caption{Top (solid) and side (dashed) angular views of the UBVRIJHK magnitudes for the 2D
      kilonova model described in Fig.~\ref{fg1:ejecta} and Section~\ref{sec:pulsr2d}.
      In Fig.~\ref{fg2a:pulsr2d}, the pulsar source luminosity is $2.88\times10^{43}$ erg/s.
      In Fig.~\ref{fg2b:pulsr2d}, the pulsar source luminosity is $2.88\times10^{45}$ erg/s.}
    \label{fg2:pulsr2d}
\end{figure*}

Strong viewing angle dependence is also seen in broadband magnitudes.
Figure~\ref{fg2:pulsr2d} has plots of UBVRIJHK absolute magnitudes for
each model.
The viewing angle dependence is stronger for bluer bands, as expected.
Before 2 days, the side view (dashed) is not sensitive to the strength of the pulsar
luminosity.

\subsection{Fallback Powered Light Curves}

Our fallback models include a range of accretion rates, $\dot{M}$ and
ejecta masses.
The variations in $\dot{M}$ bound the typical value adopted by
\cite{li2018}.
These variations in $M_{\rm ej}$ and $\dot{M}$ demonstrate the competition
between optical depth and remnant source power, similar to the pulsar variations.
Following Section~\ref{sec:pulsar_results}, we include both 1- and 2-dimensional
geometries.

\subsubsection{1-dimensional models, Accretion Rates and Ejecta Masses}
\label{sec:fvar}

For the 1D fallback models, we vary $v_{\max}$, $\dot{M}$ and $M_{\rm ej}$;
otherwise the model parameters of~\cite{li2018} for the fallback luminosity source are
adopted, as shown in Eq.~\eqref{eq1:flum}.
The ejecta masses, velocity, and composition are the same here as in
Section~\ref{sec:vars}, permitting direct comparison between these fallback
models and a subset of the pulsar models.
The model parameters are given in Table~\ref{tb1:fvar}
\begin{table}[H]
  \centering
  \caption{Parameters for test of mass and accretion rate variation.}
  \label{tb1:fvar}
  \resizebox{0.75\columnwidth}{!}{
  \begin{tabular}{c|c}
    \hline
    $L_0$ & $2\times10^{51}$ erg/s \\
    $t_{\rm acc}$ & 0.1 s \\
    $M_{\rm ej}$ & $\{1, 3, 10\}\times10^{-3}$ M$_{\odot}$ \\
    $v_{\max}/2$ & $\{0.3, 0.45\}\,c$ \\
    $\eta$ & $0.1$ \\
    $\dot{M}$ & $\{1, 3, 10\}\times10^{-3}$ M$_{\odot}$/s \\
    $(\chi_{\rm Fe}, \chi_{\rm Nd})$ & $(1-10^{-4}, 10^{-4})$ \\
    \hline
  \end{tabular}
  }
\end{table}

\begin{figure}
    \centering
    \subfloat[]{\includegraphics[width=0.5\textwidth]{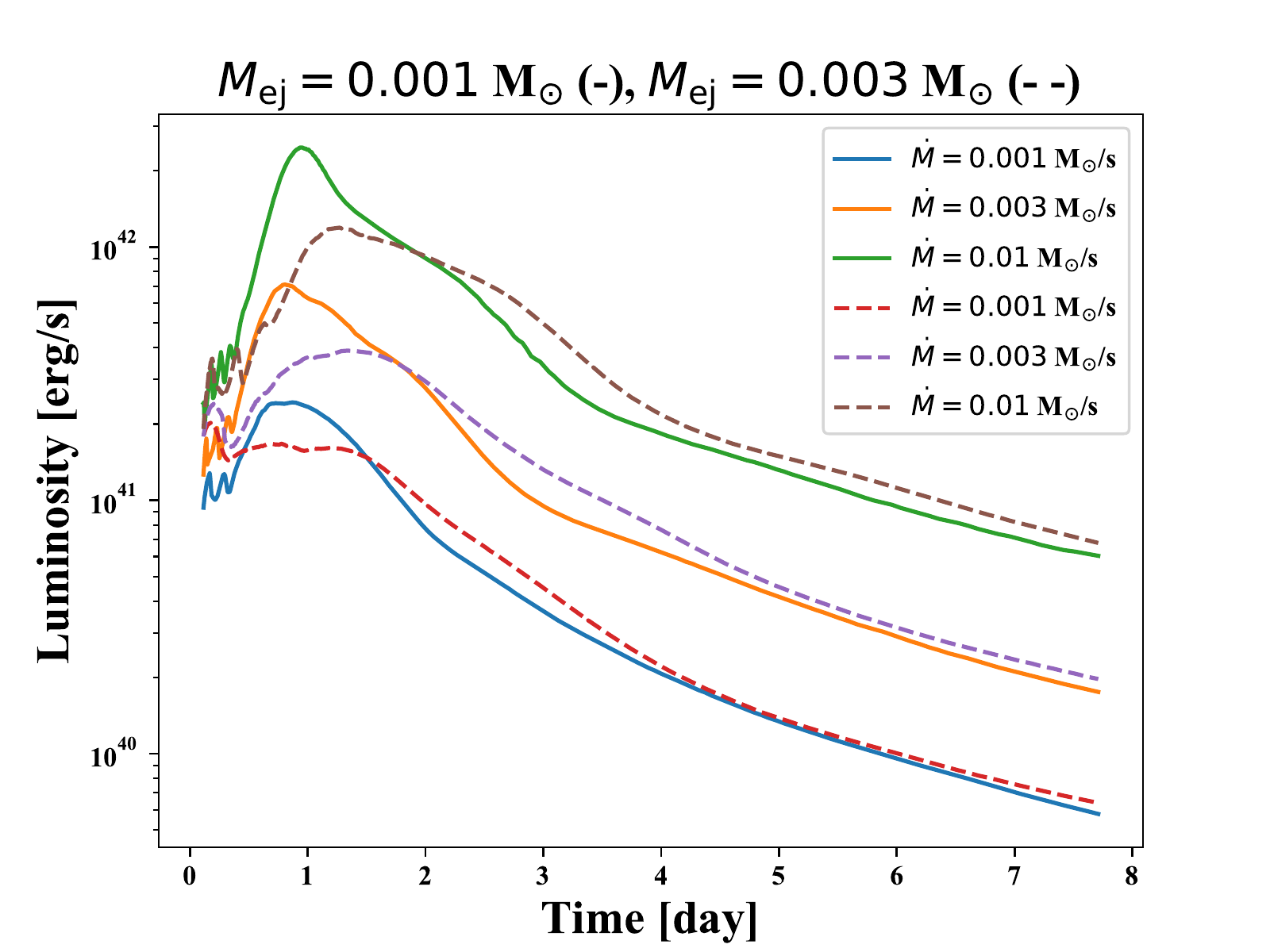}}
    \caption{Bolometric luminosity versus time for the fallback kilonova model described
      in Section~\ref{sec:fvar}.
      The light curves were obtained under the assumption of an ejecta mass of
      0.001 M$_{\odot}$ (solid) or 0.003 M$_{\odot}$ (dashed) and the three accretion rates.
    }
    \label{fg1:fvar}
\end{figure}

Figure~\ref{fg1:fvar} displays bolometric luminosity versus time for the different
accretion rates.
Increasing the ejecta mass acts to dim the light curve, since the fallback source
is more obscured at higher optical depth.
This indicates that much of the kilonova luminosity in these models is derived
from radiated accretion energy.

In the Appendix, UBVRIJHK band data at day 1 and day 7 are provided
for these fallback models in Tables~\ref{tb1:ftables}--\ref{tb2:ftables}.
The accretion rate impacts the rate of decline in the optical bands. For instance,
for $M_{\rm ej}=0.001$ M$_{\odot}$ and $v_{\max}/2=0.3c$, accretion rates of
$\dot{M}=0.001$, $\dot{M}=0.003$, and $\dot{M}=0.01$ M$_{\odot}$/s drop by $\sim$4.2,
3.5, and 2.8 magnitudes in the V-band, respectively.
Likewise, the R-band drops by $\sim$3.9, 3.2, and 2.7 magnitudes, respectively.
The J through K-bands are relatively insensitive to the fallback source at
day 1, but their brightness is more affected by day 7, where the variation with
accretion rate is $> 1$ magnitude.
In other words, at early time, the JKH magnitudes appear to be mainly set by
$M_{\rm ej}$, while at later time, there is a systematic increase in magnitude
with increasing accretion rate.
For the high-velocity models, day 1 is later in the evolution of the light curves,
so these trends are not reflected in the broadband data.

\subsubsection{2-component, 2-dimensional Accretion Results}
\label{sec:fallb2d}

\begin{figure}
  \centering
  \includegraphics[width=0.45\textwidth]{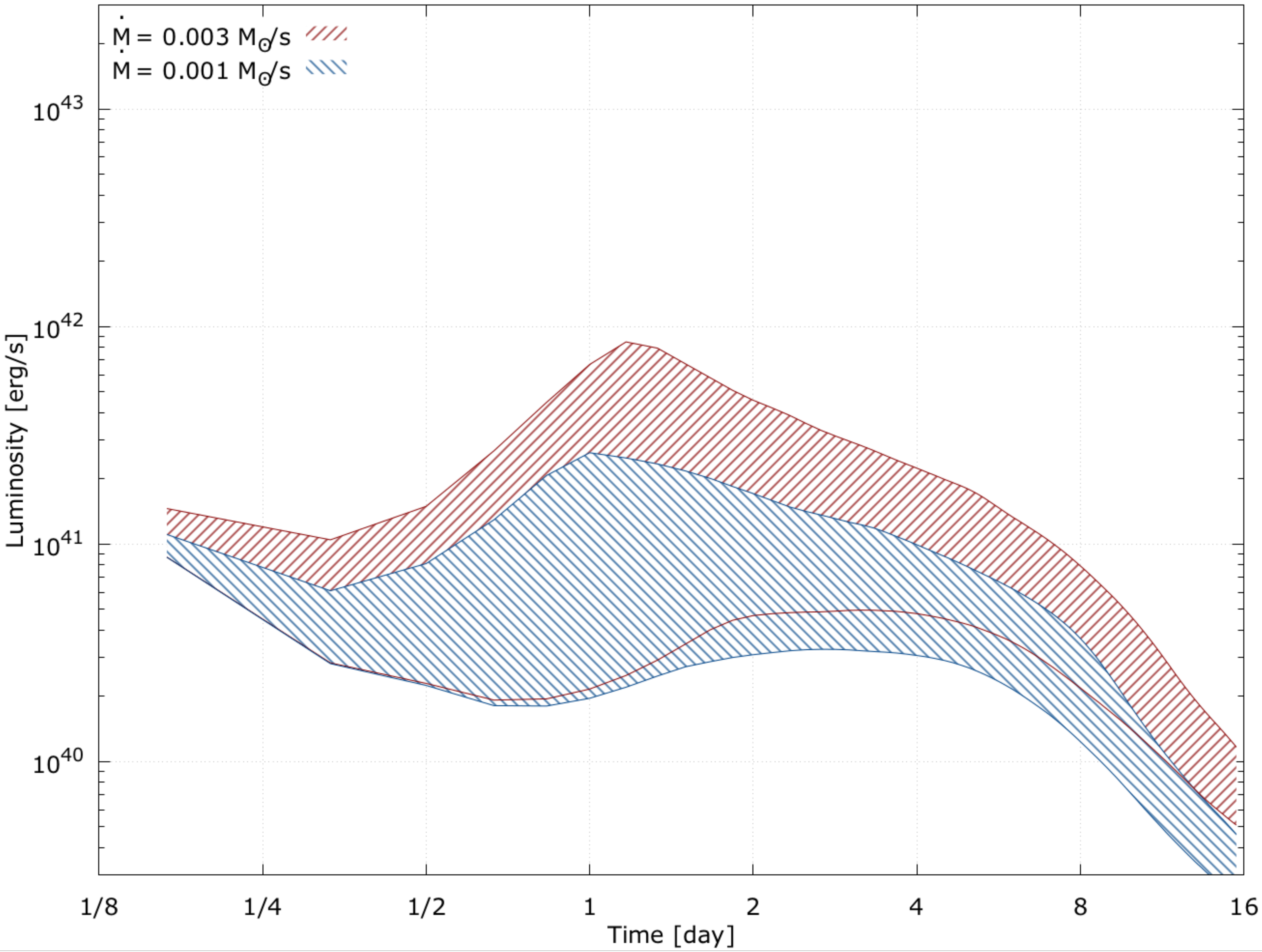}
  \caption{Isotropically equivalent bolometric luminosity versus time for the
    2D fallback kilonova models
    described in Fig.~\ref{fg1:ejecta} and Section~\ref{sec:fallb2d}.
    The shaded regions represent the range of luminosities from the 54 angular
    bins that are uniform in the cosine of the polar viewing angle.
    Brighter luminosity corresponds to viewing bins that are more aligned with
    the merger axis.
  }
  \label{fg1:fallb2d}
\end{figure}

For the fallback models in 2D, we again use the ``model A'' morphology, as in
Section~\ref{sec:pulsr2d}.
The wind superimposed on the model A ejecta has the properties described
in Section~\ref{sec:fvar} for $\dot{M}=0.001$ and 0.003 M$_{\odot}$/s, and with a wind
mass of 0.001 M$_{\odot}$.
Figure~\ref{fg1:fallb2d} has bolometric luminosity for the range of angular views
as shaded regions for each model.
As in Fig.~\ref{fg1:pulsr2d}, the values are shown as isotropic equivalents.
The angular views are again divided into 54 polar angular viewing ranges of
equal solid angle, with edge-on views being dimmest.
Relative to the 2D pulsar models, the variation with respect to viewing angle does
not change as substantially when increasing the source luminosity.
This is partly explained by the fallback luminosity scaling linearly with accretion
rate, whereas the pulsar source luminosity scales as the square of the magnetic
field strength.
As in the 2D pulsar models, increasing the remnant source luminosity affects the
rise and decline times for near-on-axis angular views of the ejecta.

\begin{figure*}
    \centering
    \subfloat[]{\includegraphics[width=0.5\textwidth]{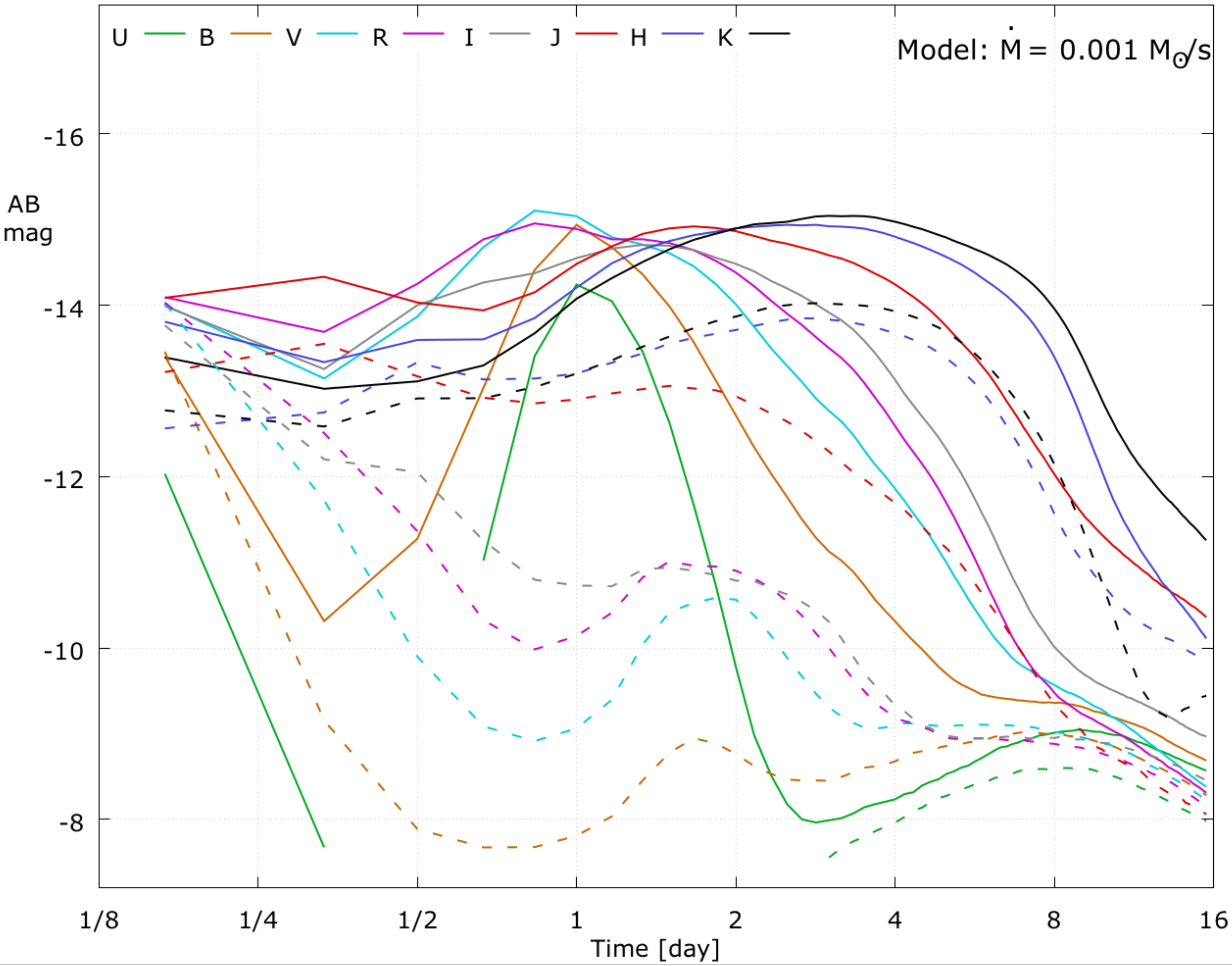}\label{fg2a:fallb2d}}
    \subfloat[]{\includegraphics[width=0.5\textwidth]{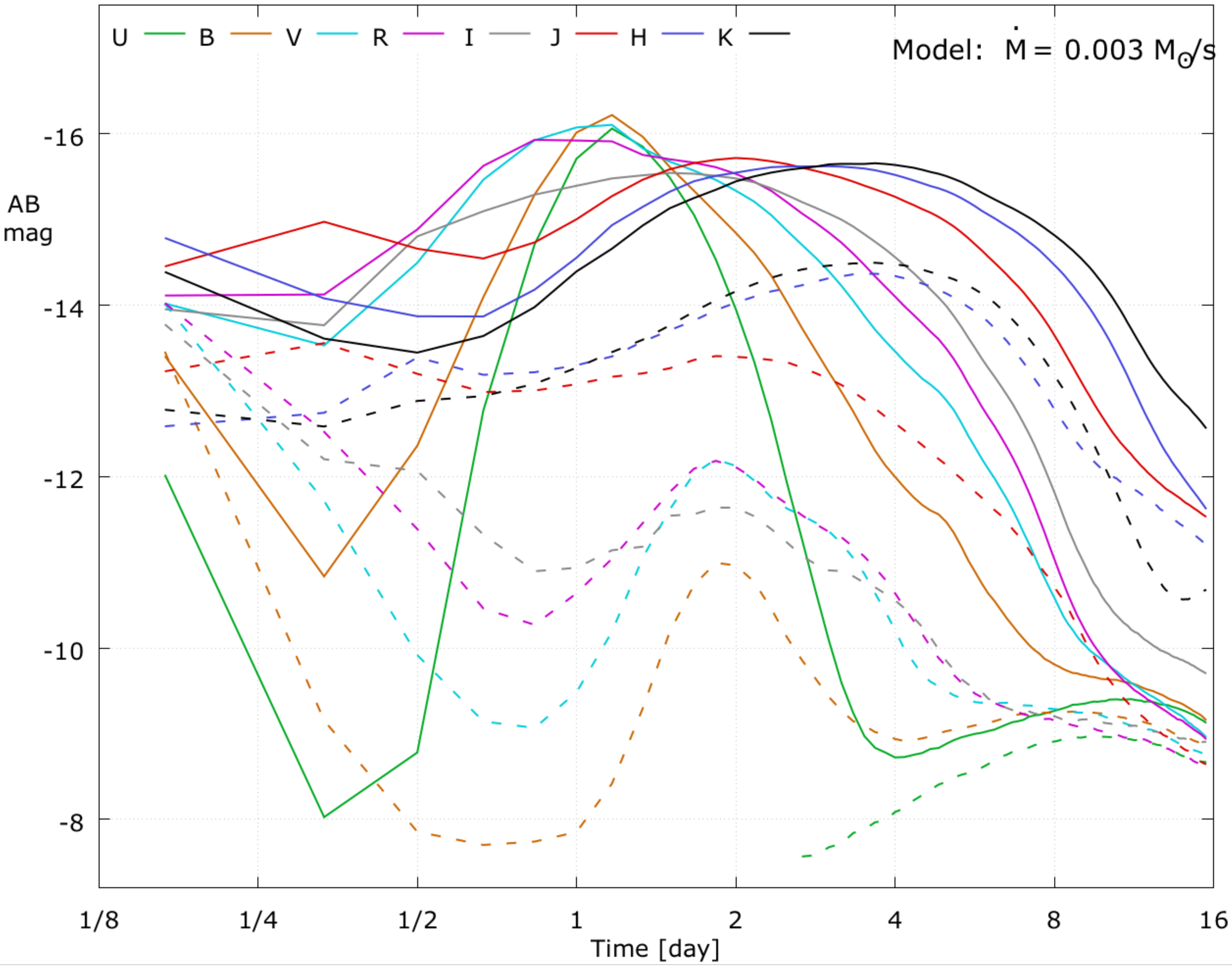}\label{fg2b:fallb2d}}
    \caption{Top (solid) and side (dashed) angular views of the UBVRIJHK magnitudes for the 2D
      kilonova model described in Fig.~\ref{fg1:ejecta} and Section~\ref{sec:fallb2d}.
      In Fig.~\ref{fg2a:fallb2d}, the fallback accretion rate is 0.001 M$_{\odot}$/s.
      In Fig.~\ref{fg2b:fallb2d}, the fallback accretion rate is 0.003 M$_{\odot}$/s.}
    \label{fg2:fallb2d}
\end{figure*}

Strong viewing angle dependence is again seen in broadband magnitudes.
Figure~\ref{fg2:fallb2d} has plots of UBVRIJHK absolute magnitudes for
each model.
The viewing angle dependence is stronger for bluer bands, as expected.
Before 2 days, the side view (dashed) is not sensitive to the strength of the pulsar
luminosity.
Unlike the 2D pulsar models, the U-band does not significantly shift in time and
peak brightness when going to a higher remnant source, which is due to the more
modest increase in luminosity relative to the pulsar models.

\subsection{Accretion versus Pulsar Power}

\begin{figure}
    \centering
    \subfloat[]{\includegraphics[width=0.5\textwidth]{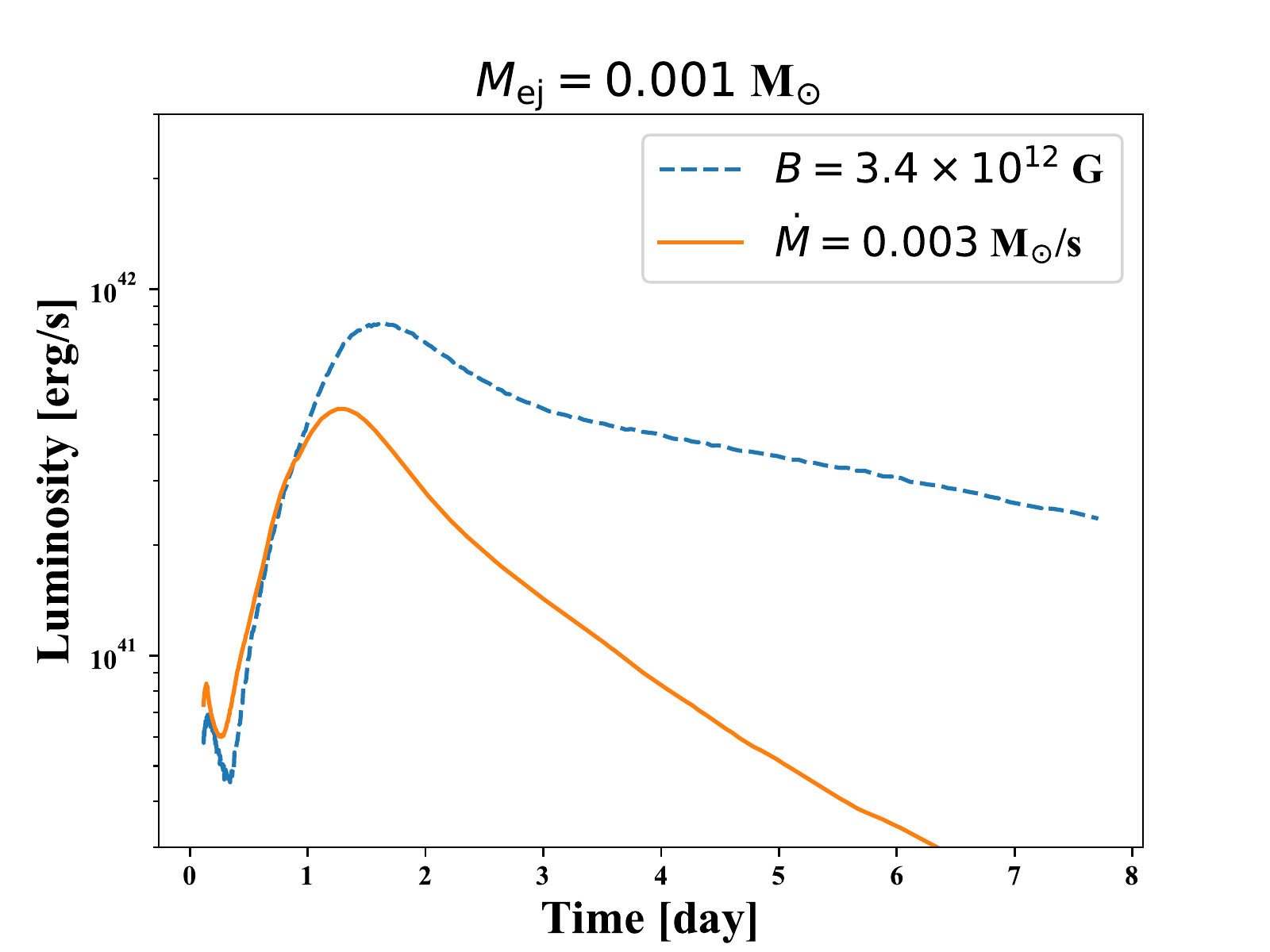}}
    \caption{Bolometric luminosity versus time comparing the fallback energy source to
      the pulsar energy source.
      The figure compares a fallback model with a $\dot{M}=0.003$ M$_{\odot}$/s to
      a pulsar model with $B=3.4\times10^{12}$ G. The slopes of the light-curve tails
      are eventually set by the time-dependence of the remnant source.}
    \label{fg1:comp}
\end{figure}

With our two simple prescriptions for accretion and pulsar power sources,
we produce different light curves.
In Figure~\ref{fg1:comp}, we compare a fallback model with $\dot{M}=0.003$
M$_{\odot}$/s to a pulsar model with $B=3.4\times10^{12}$ G. It is evident that the slope of the
light-curve tails are eventually set by the time-dependence of the remnant source.

\subsubsection{AT 2017gfo}
\label{sec:gw17}

We can compare our two different additional power sources to observations
of AT 2017gfo using the results of~\cite{li2018} and~\cite{piro2018} as a
starting point.
Although there is a qualitative agreement with those results, there are
some differences in our conclusions.
For the polar ejecta mass of 0.001 M$_{\odot}$ given by~\cite{li2018},
our ejecta model requires a slightly lower pulsar luminosity and somewhat
higher wind outflow speed than previously published: $2\times10^{44}$ instead
of $3.4\times10^{44}$ erg/s and $0.45c$ instead of $0.35c$.
Additionally, we find that this model fits the later time bolometric
luminosity better with a remnant cutoff time of about $\sim10$ observer days.
From the modeling perspective, some differences here include the use of
multifrequency opacity tables and a detailed high-latitude r-process heating
rate (corresponding to the ``Wind 1'' composition of~\cite{wollaeger2018})
derived from {\tt WinNet}~\citep{winteler2012a,winteler2012b}.
For the fallback model, similar to the polar model of~\cite{li2018}, we find
an accretion rate of 0.003-0.0045 M$_{\odot}$/s gives bolometric luminosity in
the range of the observation, but with a velocity of $\sim0.4c$ instead of $0.25c$.
Parameters for the pulsar and fallback models are given in Tables and, respectively.

\begin{table}[H]
  \centering
  \caption{Parameters for pulsar fit of AT 2017gfo.}
  \label{tb1:gw17}
  \resizebox{0.65\columnwidth}{!}{
  \begin{tabular}{c|c}
    \hline
    $M_{\rm ej}$ & $10^{-3}$ M$_{\odot}$ \\
    $v_{\max}/2$ & $0.45\,c$ \\
    $L_0$ & $2\times10^{44}$ erg/s \\
    $t_{\rm gw}$ & 495 s \\
    $t_{\rm cut}$ & $2\times10^6$ s \\
    $(\chi_{\rm Fe}, \chi_{\rm Nd})$ & $(1-10^{-3}, 10^{-3})$ \\
    \hline
  \end{tabular}
  }
\end{table}

\begin{table}[H]
  \centering
  \caption{Parameters for fallback fit of AT 2017gfo.}
  \label{tb2:gw17}
  \resizebox{0.65\columnwidth}{!}{
  \begin{tabular}{c|c}
    \hline
    $M_{\rm ej}$ & $10^{-3}$ M$_{\odot}$ \\
    $v_{\max}/2$ & $0.4\,c$ \\
    $L_0$ & $2\times10^{51}$ erg/s \\
    $\eta$ & $0.1$ \\
    $\dot{M}$ & $4.5\times10^{-3}$ M$_{\odot}$/s \\
    $t_{\rm acc}$ & 0.1 s \\
    $(\chi_{\rm Fe}, \chi_{\rm Nd})$ & $(1-10^{-3}, 10^{-3})$ \\
    \hline
  \end{tabular}
  }
\end{table}

\begin{figure*}
  \centering
  \subfloat[Pulsar]{\includegraphics[width=0.5\textwidth]{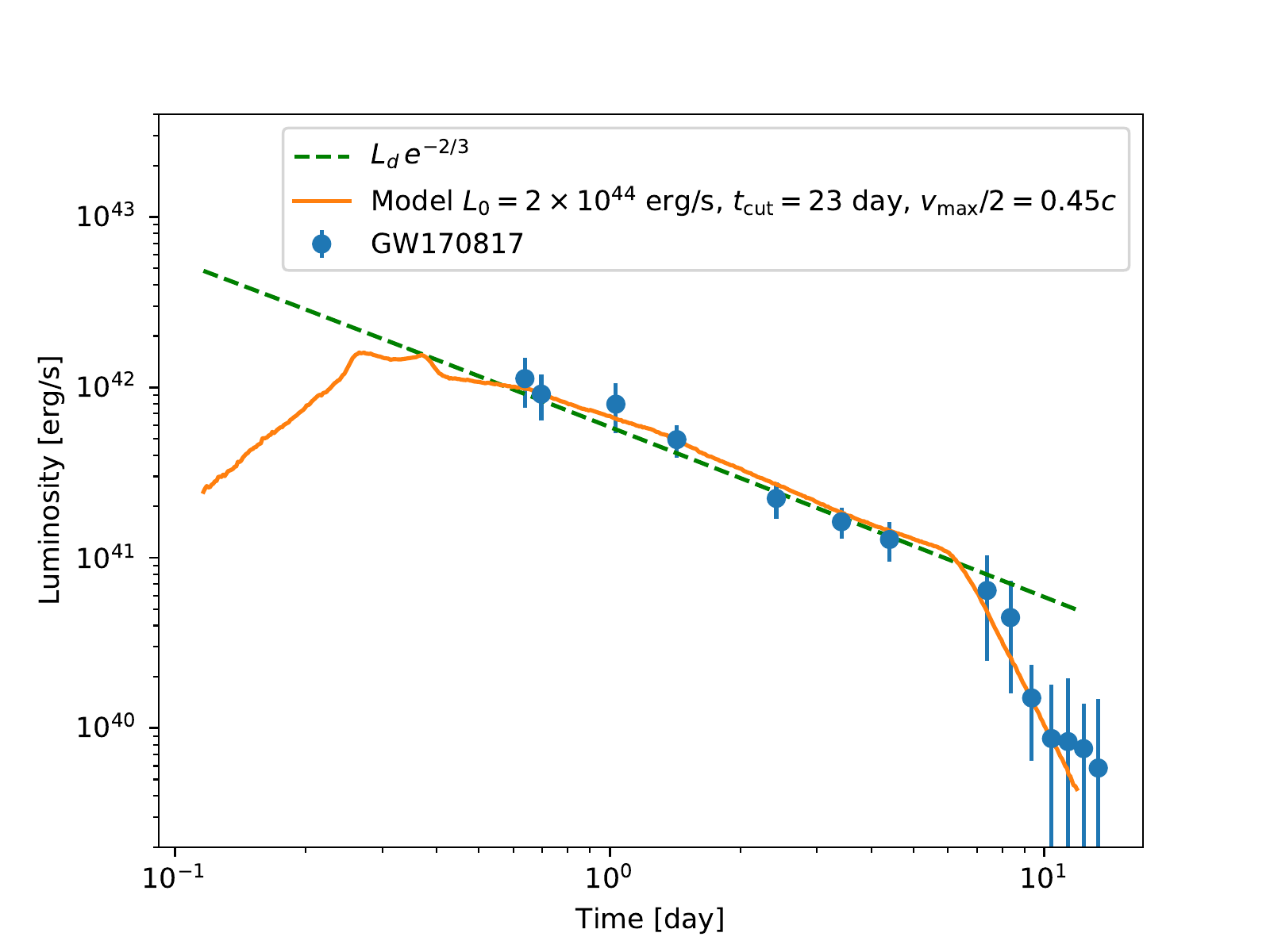}\label{fg1a:gw17}}
  \subfloat[Pulsar]{\includegraphics[width=0.5\textwidth]{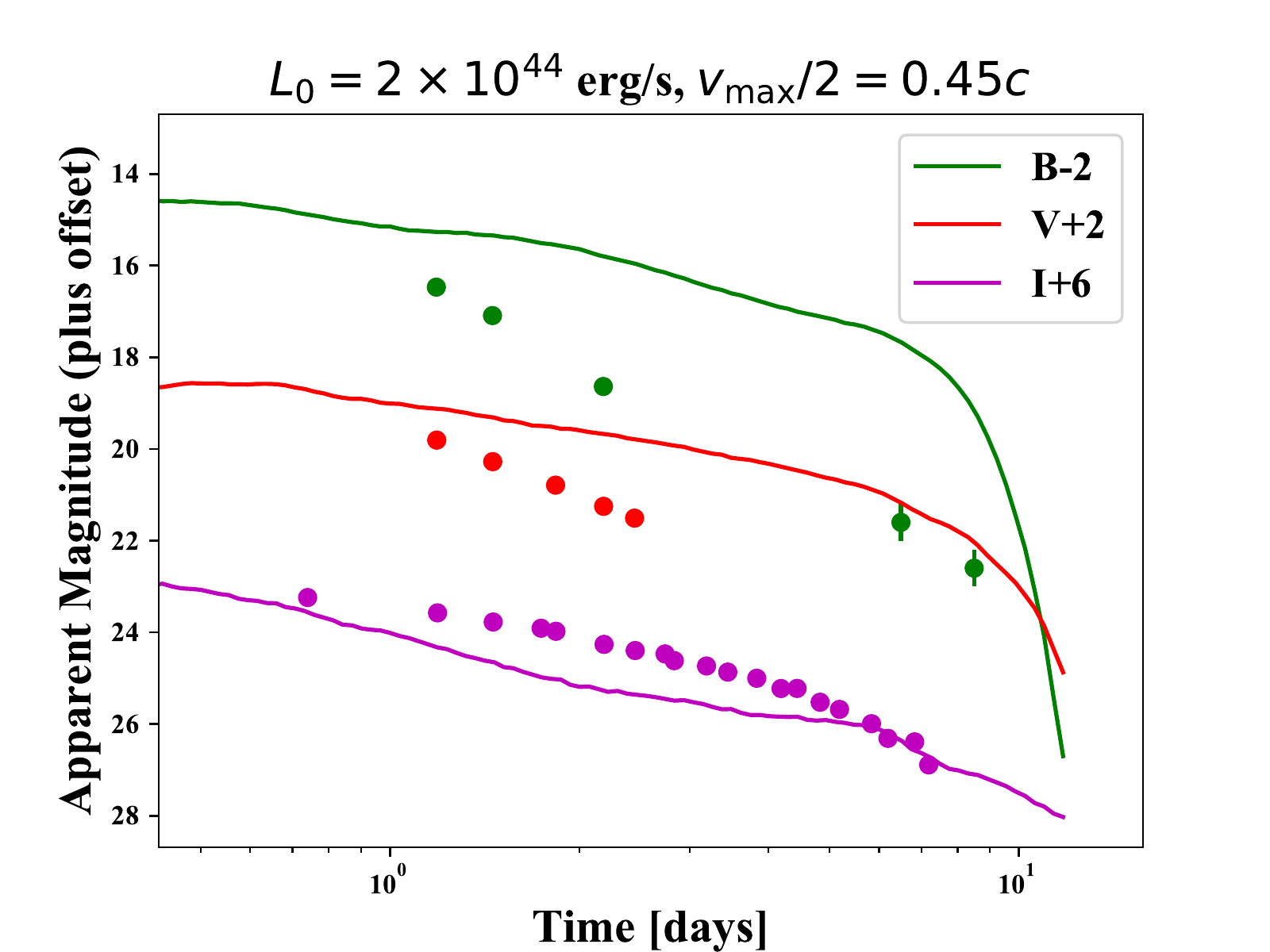}\label{fg1b:gw17}} \\
  \subfloat[Fallback]{\includegraphics[width=0.5\textwidth]{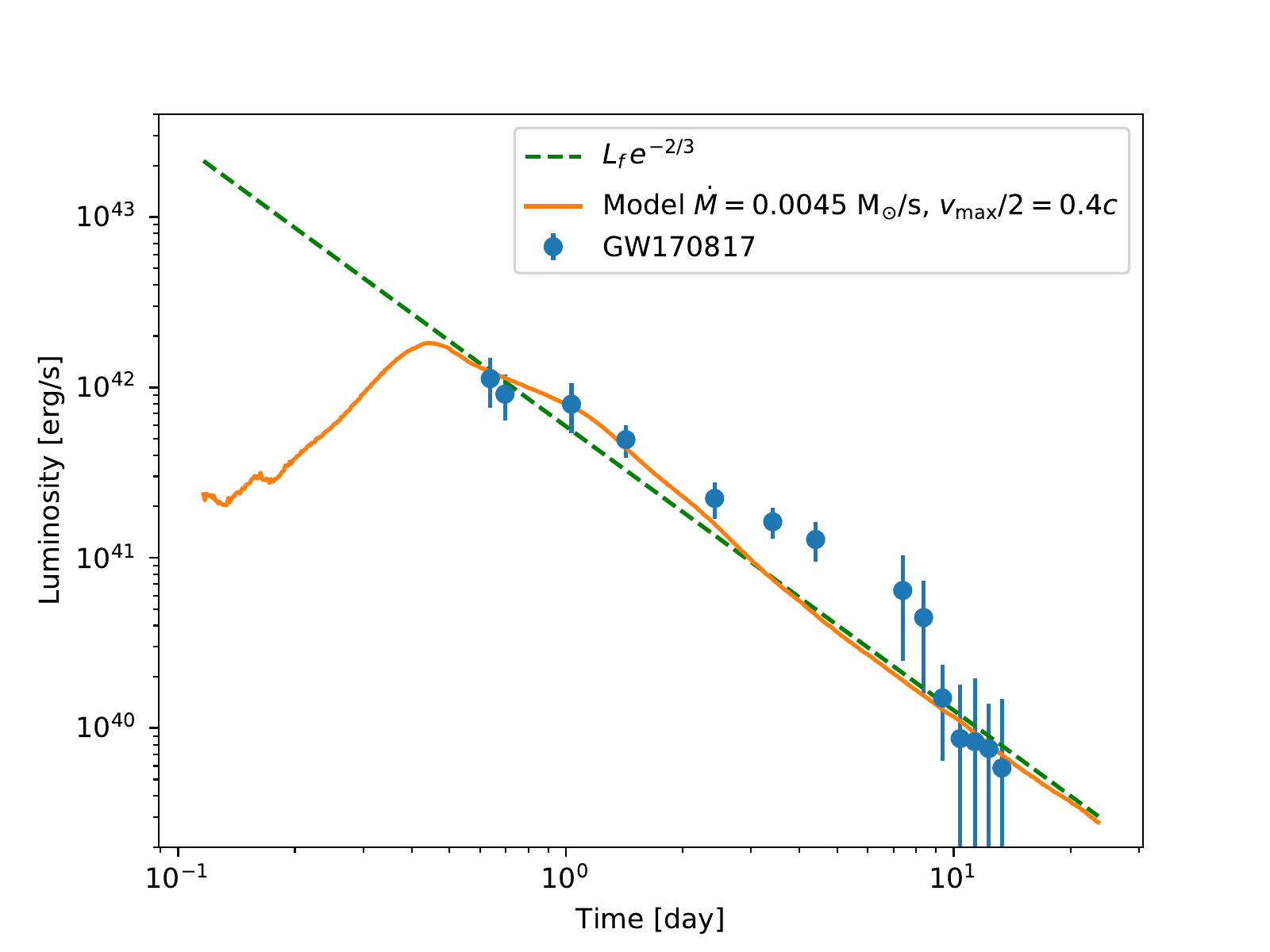}\label{fg1c:gw17}}
  \subfloat[Fallback]{\includegraphics[width=0.5\textwidth]{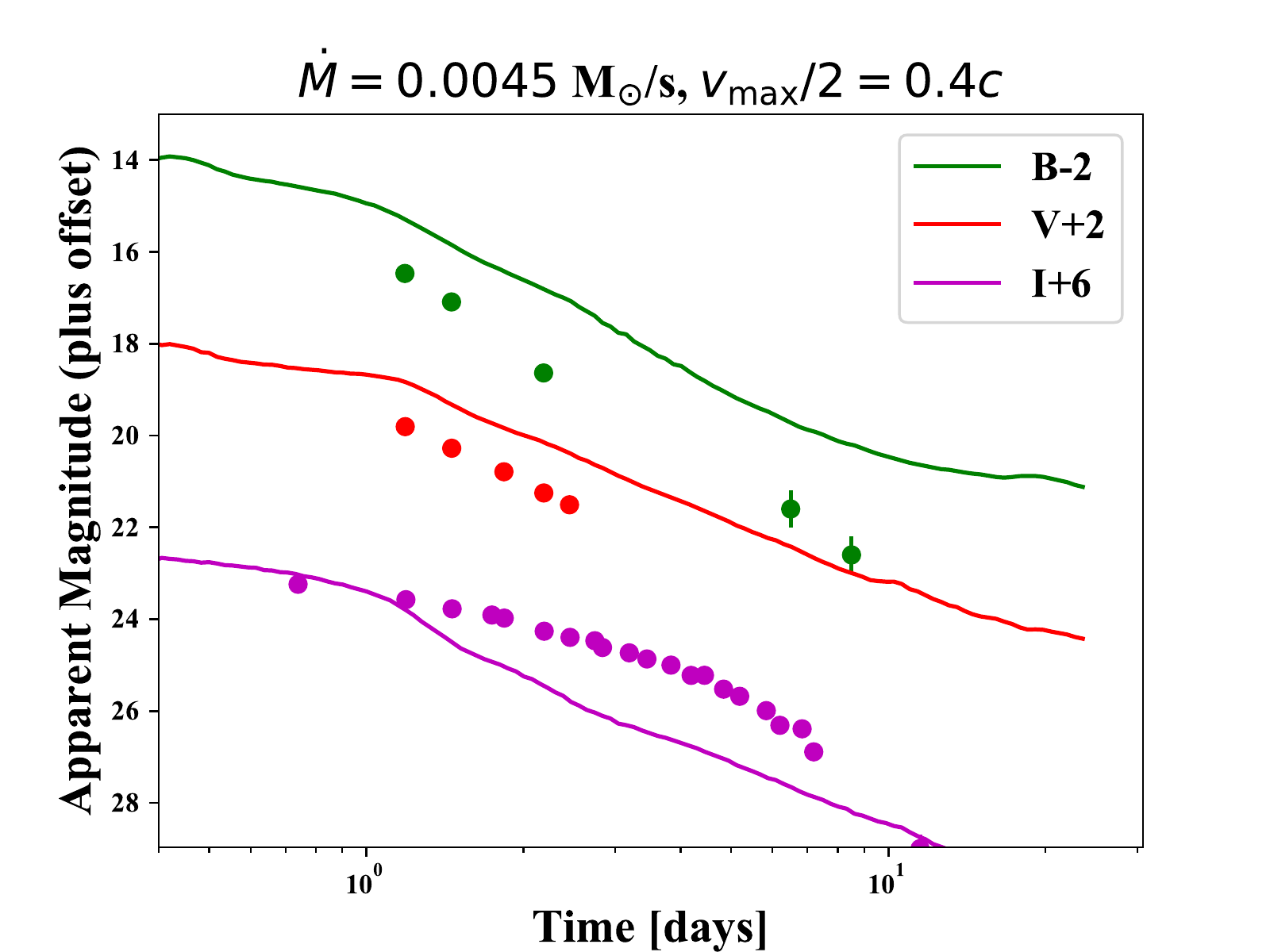}\label{fg1d:gw17}}
  \caption{
    In Figs.~\ref{fg1a:gw17} and~\ref{fg1c:gw17}, bolometric luminosity versus time
    of the pulsar and fallback source models, respectively, described in Section~\ref{sec:gw17},
    along with AT 2017gfo luminosity data from~\cite{piro2018}.
    Overplotted are the remnant source luminosities (dashed green),
    Eqs.~\eqref{eq4:plum} and~\eqref{eq1:flum} multiplied by an attenuation factor,
    $e^{-2/3}\approx 0.5$.
    In Figs.~\ref{fg1b:gw17} and~\ref{fg1d:gw17}, B, V, and I broadband magnitude
    versus time for the pulsar and fallback source models in Section~\ref{sec:gw17}
    along with broadband data from~\cite{troja2017}.
    The trends in the B and V-bands of the observed data are better fit by the fallback
    model, which has a source luminosity that declines more rapidly (Eq.~\eqref{eq1:flum}).
    Figures~\ref{fg1a:gw17} and~\ref{fg1b:gw17} are derived from the spectra of the
    model presented in Table~\ref{tb1:gw17}, and Figs.~\ref{fg1c:gw17} and~\ref{fg1d:gw17}
    are from the spectra of the model presented in Table~\ref{tb2:gw17}.
  }
  \label{fg1:gw17}
\end{figure*}

Figure~\ref{fg1:gw17} has bolometric and broadband luminosity versus time for
our attempt to match AT 2017gfo with a pulsar and a fallback model.
Despite the decent agreement in bolometric luminosity in Fig.~\ref{fg1a:gw17},
it is apparent in~\ref{fg1b:gw17} that this model produces blue emission that does
not decay quickly enough with respect to the observation.
Figure~\ref{fg1c:gw17} has the light curve for the fallback model.
The fallback model bolometric luminosity does not fit the data as well at
intermediate times, implying the decline in the source luminosity
is too rapid to fully account for this emission, as found by~\cite{li2018}.
However, the trends in the B and V bands are much more similar to those of
AT 2017gfo, relative to the pulsar model.
The I-band of the fallback model is not bright enough at later times, which
may be part of the disk.

\section{Conclusions}

We have explored several models of remnant source energy contributions
to the kilonova optical/IR signal.
These models are to serve as part of a larger database of models to which
future observed kilonovae can be compared.
Below we summarize our findings for this subset of models.
\begin{itemize}
\item Our attempt to fit AT 2017gfo in both bolometric luminosity and
  broadband light curves is unsuccessful.
  For a pulsar model fit of the bolometric luminosity, the broadband data
  is too blue in our tests, despite having an I-band in reasonable agreement
  with the observation.
  On the other hand, for the fallback model fit, the B and V-bands appear
  to be more consistent, but the I-band declines too rapidly.
  Moreover, the trend in the bolometric luminosity of the fallback model
  does not appear to fit well from a few days to a week.
  However, we stress that our attempt to fit AT 2017gfo is not an exhaustive
  study, and that caution must be taken when attempting to draw conclusions
  from the fits.
  In particular, while the calculations were multi-frequency, we only tested
  with Fe and Nd.
  Expanding the set of elements in these models could affect the trends in
  the broadband magnitudes.
\item For the grid of models in which the remnant cutoff time, $t_{\rm cut}$, was varied,
  we find that, for a pulsar luminosity comparable to that of~\cite{li2018},
  $t_{\rm cut} \gtrsim$ 1 day saturates the peak luminosity.
  The velocity of $\sim0.3c$ evidently permits the effective photosphere
  to recede quickly enough that the pulsar energy can be uncovered on
  a $\gtrsim 1$ day timescale.
\item As expected, increasing the pulsar luminosity and lifetime generally
  increases the brightness of these models.
  Increasing ejecta mass may either increase or decrease the kilonova
  luminosity.
  In particular, going from $0.001$ to $0.01$ M$_{\odot}$ increases the r-process
  heating enough to mask the contribution of the low pulsar luminosity
  ($\sim10^{43}$ erg/s) to the kilonova luminosity at 1 day.
  In contrast, with high pulsar luminosity, making the same increases to the
  ejecta mass lowers the luminosity at 1 day.
  This behavior is due to the optical depth increasing while the total r-process heating
  remains sub-dominant with respect to the pulsar luminosity.
\item With a small contribution of Nd, the H and K bands remain
  brighter at day 7 for models with ejecta mass $\geq0.003$ M$_{\odot}$
  and magnetic field $\geq10^{13}$ G, despite these models having similar
  broadband luminosity at day 1 to the versions with pure Fe.
\item Following~\cite{metzger2017,li2018}, the luminosity from fallback can
  significantly impact the luminosity of the kilonova model.
  As expected, the decay of the kilonova bolometric light-curve tail is set by
  the time-dependence of the remnant source.
  For similar source luminosity, the spectra from the fallback models are
  similar to those of the pulsar models; it is primarily a function of the
  ejecta composition.
  The slower decline of the luminosity tails of the pulsar models permit
  them to stay more luminous and have bluer spectra at later time.
\item The 2D models show that the remnant source luminosity can affect
  viewing-angle dependence of the light curves.
  In particular, at early time (for instance, $\lesssim1/2$ day for the pulsar
  models explored), edge-on (off-axis) views of the ejecta produce broadband
  and bolometric light curves that are insensitive to the brightness of the
  pulsar source.
  As the photosphere recedes through the dynamical ejecta, the remnant source
  begins to contribute to the observable light curve.
  After 1/2 day, the time of peak magnitude for the U and B bands is
  shifted earlier by a few days in the higher-source energy model, while the edge
  on views for these bands are substantially re-brightened after the initial time.
\end{itemize}

Future work will involve attempting to refine the parameters, identifying model
deficiencies, and comparing to new observations during the LIGO O3 run.
The models presented in this work are a first step towards these efforts.
Some concepts we have gleaned from the remnant source study, that may be useful
to further modeling efforts to match AT 2017gfo, are (1) the dependence of
bolometric luminosity and broadband magnitudes on the time dependence of the
remnant source, and (2) the competition between r-process decay heating, the
remnant source luminosity, and optical depth for certain model parameters.

\section*{Acknowledgments}

This work was supported by the US Department of Energy through the Los Alamos
National Laboratory. Los Alamos National Laboratory is operated by Triad
National Security, LLC, for the National Nuclear Security Administration
of U.S.\ Department of Energy (Contract No.\ 89233218CNA000001).
Research presented in this article was supported by the Laboratory Directed
Research and Development program of Los Alamos National Laboratory under
project number 20190021DR.
We thank our anonymous reviewer for the constructive input on this work.


\appendix

\section{Luminosity \& Broadband Magnitudes for Pulsar Kilonova Models}
\label{sec:ptables}

\begin{table}[H]
  \centering
  \caption{Luminosity (erg/s) and 40 Mpc UBVRIJHK bands at day 1 of SAFe model
    from Section~\ref{sec:vars}.}
  \label{tb1:tables}
  \resizebox{0.8\columnwidth}{!}{
  \begin{tabular}{|r|ccccccccc|}
    \hline
    $t_{\rm cut}$ (s), $B$ (G), $M_{\rm ej}$ (M$_{\odot}$) & $L_{\rm bol}$ &
    U & B & V & R & I & J & H & K \\
    \hline
    $2\times10^4$, $10^{12}$, 0.001 & $1.41\times10^{40}$
    & 30.1 & 27.0 & 23.3 & 21.6 & 21.6 & 19.7 & 20.0 & 23.4 \\
    0.003 & $1.74\times10^{40}$
    & 29.4 & 26.5 & 23.8 & 21.4 & 21.0 & 19.7 & 19.8 & 21.5 \\
    0.01 & $5.17\times10^{40}$
    & 28.1 & 25.3 & 22.9 & 20.5 & 19.8 & 18.9 & 18.5 & 19.3 \\
    $10^{13}$, 0.001 & $7.13\times10^{41}$
    & 17.0 & 16.8 & 16.5 & 17.0 & 18.2 & 18.8 & 19.1 & 21.7 \\
    0.003 & $2.60\times10^{41}$
    & 24.7 & 21.9 & 17.8 & 17.1 & 17.7 & 17.8 & 18.4 & 20.1 \\
    0.01 & $1.40\times10^{41}$
    & 27.7 & 24.6 & 21.0 & 18.4 & 18.0 & 17.8 & 17.9 & 18.9 \\
    $10^{14}$, 0.001 & $3.11\times10^{43}$
    & 13.1 & 14.0 & 12.5 & 11.9 & 13.7 & 15.2 & 17.1 & 17.8 \\
    0.003 & $4.65\times10^{42}$
    & 19.5 & 17.2 & 14.8 & 14.0 & 14.7 & 14.6 & 16.0 & 16.7 \\
    0.01 & $1.93\times10^{42}$
    & 22.2 & 19.8 & 17.0 & 16.2 & 16.7 & 14.7 & 15.1 & 15.8 \\
    $2\times10^5$, $10^{12}$, 0.001 & $9.96\times10^{40}$
    & 22.7 & 20.2 & 18.2 & 18.4 & 20.2 & 19.1 & 19.5 & 23.3 \\
    0.003 & $6.48\times10^{40}$
    & 28.4 & 25.8 & 20.5 & 18.8 & 19.1 & 18.9 & 19.2 & 21.1 \\
    0.01 & $8.48\times10^{40}$
    & 27.6 & 25.0 & 22.0 & 19.2 & 18.6 & 18.5 & 18.3 & 19.1 \\
    $10^{13}$, 0.001 & $4.55\times10^{42}$
    & 15.1 & 15.6 & 14.6 & 14.1 & 15.8 & 17.4 & 18.9 & 19.9 \\
    0.003 & $2.34\times10^{42}$
    & 17.4 & 16.6 & 15.2 & 14.9 & 15.5 & 16.4 & 17.7 & 18.8 \\
    0.01 & $5.36\times10^{41}$
    & 24.4 & 21.9 & 17.9 & 16.9 & 17.0 & 16.1 & 16.8 & 17.7 \\
    $10^{14}$, 0.001 & $8.52\times10^{43}$
    & 12.5 & 13.5 & 11.6 & 10.8 & 11.5 & 12.7 & 14.5 & 15.3 \\
    0.003 & $3.67\times10^{43}$
    & 15.8 & 15.6 & 13.1 & 12.1 & 12.2 & 12.1 & 13.5 & 14.2 \\
    0.01 & $1.91\times10^{43}$
    & 19.2 & 17.2 & 15.6 & 14.6 & 14.1 & 12.2 & 12.6 & 13.2 \\
    \hline
  \end{tabular}
  }
\end{table}

\begin{table}[H]
  \centering
  \caption{Luminosity (erg/s) and 40 Mpc UBVRIJHK bands at day 1 of SAFeNd model
    from Section~\ref{sec:vars}.}
  \label{tb2:tables}
  \resizebox{0.8\columnwidth}{!}{
  \begin{tabular}{|r|ccccccccc|}
    \hline
    $t_{\rm cut}$ (s), $B$ (G), $M_{\rm ej}$ (M$_{\odot}$) & $L_{\rm bol}$ &
    U & B & V & R & I & J & H & K \\
    \hline
    $2\times10^4$, $10^{12}$, 0.001 & $1.49\times10^{40}$
    & 30.6 & 27.4 & 23.9 & 21.8 & 21.5 & 19.6 & 20.0 & 22.2 \\
    0.003 & $1.88\times10^{40}$
    & 29.6 & 26.7 & 23.9 & 21.4 & 21.0 & 19.6 & 19.7 & 21.0 \\
    0.01 & $5.37\times10^{40}$
    & 28.3 & 25.5 & 22.9 & 20.6 & 19.8 & 18.8 & 18.5 & 19.0  \\
    $10^{13}$, 0.001 & $7.02\times10^{41}$
    & 17.1 & 16.8 & 16.5 & 17.0 & 18.1 & 18.3 & 19.0 & 21.1 \\
    0.003 & $2.65\times10^{41}$
    & 24.9 & 22.1 & 17.9 & 17.1 & 17.7 & 17.6 & 18.4 & 19.7  \\
    0.01 & $1.39\times10^{41}$
    & 27.7 & 24.6 & 21.0 & 18.5 & 18.1 & 17.8 & 18.0 & 18.6  \\
    $10^{14}$, 0.001 & $3.10\times10^{43}$
    & 13.2 & 14.0 & 12.6 & 11.9 & 13.7 & 15.2 & 16.9 & 17.9  \\
    0.003 & $4.52\times10^{42}$
    & 19.2 & 17.1 & 14.8 & 14.1 & 14.8 & 14.6 & 16.0 & 16.8  \\
    0.01 & $1.76\times10^{42}$
    & 22.5 & 19.8 & 16.9 & 16.2 & 16.8 & 14.6 & 15.2 & 15.8 \\

    $2\times10^5$, $10^{12}$, 0.001 & $1.00\times10^{41}$
    & 23.5 & 20.7 & 18.4 & 18.4 & 20.0 & 18.6 & 19.4 & 21.6 \\
    0.003 & $6.86\times10^{40}$
    & 28.6 & 25.9 & 20.7 & 18.8 & 19.1 & 18.6 & 19.1 & 20.5  \\
    0.01 & $8.57\times10^{40}$
    & 27.9 & 25.1 & 22.0 & 19.3 & 18.7 & 18.4 & 18.3 & 18.9  \\
    $10^{13}$, 0.001 & $4.52\times10^{42}$
    & 15.1 & 15.6 & 14.6 & 14.1 & 15.8 & 17.3 & 18.7 & 19.7  \\
    0.003 & $2.31\times10^{42}$
    & 17.5 & 16.6 & 15.2 & 14.9 & 15.5 & 16.2 & 17.5 & 18.7  \\
    0.01 & $5.10\times10^{41}$
    & 25.2 & 22.5 & 18.1 & 16.9 & 17.1 & 16.1 & 16.9 & 17.6  \\
    $10^{14}$, 0.001 & $8.38\times10^{43}$
    & 12.5 & 13.5 & 11.6 & 10.8 & 11.5 & 12.7 & 14.4 & 15.3  \\
    0.003 & $3.63\times10^{43}$
    & 15.5 & 15.4 & 13.1 & 12.4 & 12.3 & 12.1 & 13.5 & 14.1  \\
    0.01 & $1.81\times10^{43}$
    & 18.5 & 16.6 & 15.4 & 14.6 & 14.3 & 12.2 & 12.6 & 13.2 \\
    \hline
  \end{tabular}
  }
\end{table}

\begin{table}[H]
  \centering
  \caption{Luminosity (erg/s) and 40 Mpc UBVRIJHK bands at day 7 of SAFe model
    from Section~\ref{sec:vars}.}
  \label{tb3:tables}
  \resizebox{0.8\columnwidth}{!}{
  \begin{tabular}{|r|ccccccccc|}
    \hline
    $t_{\rm cut}$ (s), $B$ (G), $M_{\rm ej}$ (M$_{\odot}$) & $L_{\rm bol}$ &
    U & B & V & R & I & J & H & K \\
    \hline
    $2\times10^4$, $10^{12}$, 0.001 & $3.42\times10^{36}$
    & 35.9 & 32.9 & 30.3 & 29.4 & 29.3 & 31.1 & 30.6 & 35.1 \\
    0.003 & $2.29\times10^{37}$
    & 34.3 & 30.9 & 28.4 & 27.6 & 27.5 & 27.8 & 27.7 & 31.8 \\
    0.01 & $1.63\times10^{38}$
    & 32.8 & 29.0 & 26.7 & 25.9 & 25.8 & 25.0 & 25.1 & 28.9 \\
    $10^{13}$, 0.001 & $4.82\times10^{36}$
    & 35.9 & 32.9 & 30.3 & 29.4 & 29.1 & 30.7 & 29.4 & 30.0 \\
    0.003 & $2.84\times10^{37}$
    & 34.3 & 30.9 & 28.4 & 27.6 & 27.5 & 27.6 & 27.3 & 28.2 \\
    0.01 & $1.78\times10^{38}$
    & 32.8 & 29.0 & 26.7 & 25.9 & 25.8 & 25.0 & 24.9 & 27.5 \\
    $10^{14}$, 0.001 & $4.05\times10^{40}$
    & 23.4 & 20.1 & 19.5 & 19.9 & 21.3 & 19.5 & 25.4 & 25.8 \\
    0.003 & $2.81\times10^{41}$
    & 20.9 & 17.9 & 17.1 & 17.4 & 18.8 & 19.9 & 23.8 & 26.2 \\
    0.01 & $1.37\times10^{42}$
    & 19.1 & 16.1 & 15.4 & 15.7 & 17.9 & 19.7 & 22.1 & 22.3 \\
    $2\times10^5$, $10^{12}$, 0.001 & $3.44\times10^{36}$
    & 35.9 & 32.9 & 30.3 & 29.4 & 29.3 & 31.1 & 30.6 & 34.2 \\
    0.003 & $2.29\times10^{37}$
    & 34.3 & 30.9 & 28.4 & 27.6 & 27.5 & 27.8 & 27.7 & 31.6 \\
    0.01 & $1.63\times10^{38}$
    & 32.8 & 29.0 & 26.7 & 25.9 & 25.8 & 25.0 & 25.1 & 28.9 \\
    $10^{13}$, 0.001 & $1.06\times10^{40}$
    & 30.9 & 25.3 & 21.9 & 20.7 & 21.0 & 21.7 & 20.2 & 26.7 \\
    0.003 & $3.55\times10^{40}$
    & 28.6 & 24.3 & 21.1 & 19.8 & 19.9 & 19.4 & 19.0 & 24.2 \\
    0.01 & $1.63\times10^{41}$
    & 26.8 & 22.7 & 20.0 & 18.6 & 18.8 & 17.0 & 17.8 & 20.0 \\
    $10^{14}$, 0.001 & $4.53\times10^{42}$
    & 15.5 & 16.4 & 17.7 & 19.7 & 21.9 & 26.5 & 33.5 & 75.7 \\
    0.003 & $1.01\times10^{43}$
    & 14.5 & 15.2 & 17.8 & 18.9 & 20.4 & 25.5 & 26.5 & 31.8 \\
    0.01 & $2.16\times10^{43}$
    & 13.3 & 14.3 & 15.3 & 15.9 & 17.3 & 21.9 & 22.8 & 27.0 \\
    \hline
  \end{tabular}
  }
\end{table}

\begin{table}[H]
  \centering
  \caption{Luminosity (erg/s) and 40 Mpc UBVRIJHK bands at day 7 of SAFeNd model
    from Section~\ref{sec:vars}.}
  \label{tb4:tables}
  \resizebox{0.8\columnwidth}{!}{
  \begin{tabular}{|r|ccccccccc|}
    \hline
    $t_{\rm cut}$ (s), $B$ (G), $M_{\rm ej}$ (M$_{\odot}$) & $L_{\rm bol}$ &
    U & B & V & R & I & J & H & K \\
    \hline
    $2\times10^4$, $10^{12}$, 0.001 & $3.40\times10^{36}$
    & - & - & 44.0 & 40.5 & 38.3 & 34.7 & 34.4 & 33.8 \\
    0.003 & $2.27\times10^{37}$
    & 131.9 & 46.8 & 38.3 & 34.6 & 32.7 & 30.4 & 30.5 & 30.6 \\
    0.01 & $1.63\times10^{38}$
    & 86.4 & 38.3 & 33.2 & 30.3 & 28.6 & 26.8 & 27.2 & 27.5 \\
    $10^{13}$, 0.001 & $4.54\times10^{36}$
    & 145.7 & 60.4 & 43.9 & 39.0 & 36.4 & 32.2 & 31.8 & 29.9 \\
    0.003 & $2.58\times10^{37}$
    & - & 47.3 & 38.1 & 34.4 & 32.0 & 29.7 & 29.3 & 29.4 \\
    0.01 & $1.80\times10^{38}$
    & - & 38.1 & 33.2 & 30.2 & 28.3 & 26.6 & 26.2 & 27.1 \\
    $10^{14}$, 0.001 & $4.56\times10^{40}$
    & 23.5 & 20.2 & 19.3 & 19.8 & 21.1 & 19.2 & 20.7 & 22.2 \\
    0.003 & $2.93\times10^{41}$
    & 21.4 & 18.2 & 17.2 & 17.5 & 18.7 & 18.0 & 19.2 & 20.8 \\
    0.01 & $1.39\times10^{42}$
    & 19.7 & 16.4 & 15.5 & 15.7 & 17.7 & 16.5 & 17.8 & 19.1 \\
    $2\times10^5$, $10^{12}$, 0.001 & $3.44\times10^{36}$
    & 137.7 & 50.9 & 44.5 & 40.7 & 38.5 & 34.7 & 34.3 & 33.8 \\
    0.003 & $2.27\times10^{37}$
    & 132.0 & 45.9 & 38.2 & 34.6 & 32.8 & 30.4 & 30.5 & 30.6 \\
    0.01 & $1.63\times10^{38}$
    & 89.3 & 38.3 & 33.3 & 30.3 & 28.6 & 26.8 & 27.2 & 27.6 \\
    $10^{13}$, 0.001 & $1.02\times10^{40}$
    & 31.3 & 25.7 & 22.2 & 20.8 & 21.1 & 21.8 & 20.1 & 23.7 \\
    0.003 & $3.53\times10^{40}$
    & 29.7 & 24.9 & 21.5 & 20.0 & 19.8 & 19.6 & 18.7 & 22.1 \\
    0.01 & $1.74\times10^{41}$
    & 28.7 & 23.8 & 20.8 & 19.0 & 18.5 & 17.1 & 17.2 & 19.5 \\
    $10^{14}$, 0.001 & $4.59\times10^{42}$
    & 15.5 & 16.4 & 17.5 & 18.9 & 19.7 & 18.6 & 20.8 & 21.7 \\
    0.003 & $1.00\times10^{43}$
    & 14.5 & 15.2 & 17.3 & 18.3 & 18.2 & 18.8 & 20.0 & 20.9 \\
    0.01 & $2.21\times10^{43}$
    & 13.3 & 14.2 & 15.1 & 15.7 & 16.4 & 17.4 & 17.9 & 19.9 \\
    \hline
  \end{tabular}
  }
\end{table}

\section{Luminosity \& Broadband Magnitudes for Fallback Kilonova Models}

\begin{table}[H]
  \centering
  \caption{Luminosity (erg/s) and 40 Mpc UBVRIJHK bands at day 1 of SAFeNd model
    with fallback source, from Section~\ref{sec:fvar}.}
  \label{tb1:ftables}
  \resizebox{0.8\columnwidth}{!}{
    \begin{tabular}{|$r|^c^c^c^c^c^c^c^c^c|}
      \hline
      $v_{\max}/2$ ($c$), $M_{\rm ej}$ (M$_{\odot}$), $\dot{M}$ (M$_{\odot}$/s) &
      $L_{\rm bol}$ & U & B & V & R & I & J & H & K \\
      \hline
      0.3, 0.001, 0.001 & $1.34\times10^{41}$ & 20.0 & 18.2 & 17.4 & 17.9 & 19.4 & 18.7 & 19.1 & 21.2 \\
      , 0.003 & $3.88\times10^{41}$ & 17.4 & 16.9 & 16.8 & 17.5 & 19.1 & 19.2 & 19.5 & 21.1 \\
      , 0.01 & $9.72\times10^{41}$ & 15.8 & 15.9 & 16.1 & 16.6 & 17.8 & 18.9 & 19.7 & 21.0 \\
      , 0.003, 0.001 & $2.56\times10^{40}$ & 24.8 & 22.4 & 18.3 & 17.8 & 19.0 & 17.9 & 18.5 & 20.2 \\
      , 0.003 & $5.75\times10^{40}$ & 21.3 & 19.0 & 16.8 & 17.1 & 18.2 & 17.6 & 18.4 & 20.1 \\
      , 0.01 & $2.53\times10^{41}$ & 18.0 & 16.8 & 16.0 & 16.4 & 17.1 & 17.2 & 18.1 & 19.6 \\
      , 0.01, 0.001 & $5.14\times10^{40}$ & 27.4 & 24.8 & 20.0 & 18.2 & 18.4 & 17.9 & 18.2 & 18.8 \\
      , 0.003 & $5.18\times10^{40}$ & 26.6 & 23.8 & 18.2 & 17.2 & 17.9 & 17.3 & 17.9 & 18.7 \\
      , 0.01 & $6.18\times10^{40}$ & 25.2 & 21.7 & 17.1 & 16.6 & 17.5 & 16.8 & 17.5 & 18.4 \\
      0.45, 0.001, 0.001 & $9.62\times10^{40}$ & 23.7 & 19.4 & 18.8 & 19.1 & 20.0 & 18.6 & 19.0 & 20.2 \\
      , 0.003 & $2.72\times10^{41}$ & 19.8 & 17.6 & 17.6 & 18.2 & 18.9 & 18.4 & 19.1 & 20.4 \\
      , 0.01 & $8.55\times10^{41}$ & 16.6 & 16.5 & 16.7 & 17.4 & 18.5 & 18.4 & 19.1 & 20.4 \\
      , 0.003, 0.001 & $1.07\times10^{41}$ & 25.8 & 20.8 & 19.1 & 19.1 & 19.6 & 17.9 & 18.0 & 20.5 \\
      , 0.003 & $3.18\times10^{41}$ & 22.4 & 18.5 & 17.5 & 17.8 & 18.8 & 17.2 & 17.6 & 18.6 \\
      , 0.01 & $1.09\times10^{42}$ & 18.1 & 16.3 & 16.0 & 16.5 & 17.5 & 17.0 & 17.8 & 18.8 \\
      , 0.01, 0.001 & $1.32\times10^{41}$ & 30.0 & 25.0 & 20.8 & 19.5 & 19.3 & 17.2 & 17.5 & 20.2 \\
      , 0.003 & $4.14\times10^{41}$ & 26.2 & 21.8 & 18.2 & 17.6 & 18.3 & 16.2 & 16.6 & 19.0 \\
      , 0.01 & $1.44\times10^{42}$ & 21.6 & 17.7 & 15.6 & 15.8 & 17.1 & 15.6 & 16.1 & 17.6 \\
      \hline
    \end{tabular}
  }
\end{table}

\begin{table}[H]
  \centering
  \caption{Luminosity (erg/s) and 40 Mpc UBVRIJHK bands at day 7 of SAFeNd model
    with fallback source, from Section~\ref{sec:fvar}.}
  \label{tb2:ftables}
  \resizebox{0.8\columnwidth}{!}{
    \begin{tabular}{|$r|^c^c^c^c^c^c^c^c^c|}
      \hline
      $v_{\max}/2$ ($c$), $M_{\rm ej}$ (M$_{\odot}$), $\dot{M}$ (M$_{\odot}$/s) &
      $L_{\rm bol}$ & U & B & V & R & I & J & H & K \\
      \hline
      0.3, 0.001, 0.001 & $7.66\times10^{39}$ & 27.2 & 23.3 & 21.6 & 21.8 & 22.4 & 20.7 & 22.2 & 23.8 \\
      , 0.003 & $2.44\times10^{40}$ & 25.3 & 21.5 & 20.3 & 20.7 & 21.3 & 19.9 & 20.6 & 23.0 \\
      , 0.01 & $8.61\times10^{40}$ & 22.6 & 19.4 & 18.9 & 19.3 & 20.5 & 19.2 & 20.0 & 22.1 \\
      , 0.003, 0.001 & $9.21\times10^{39}$ & 31.8 & 24.4 & 22.3 & 21.8 & 22.2 & 20.2 & 22.1 & 23.3 \\
      , 0.003 & $3.45\times10^{40}$ & 28.4 & 22.4 & 20.7 & 20.6 & 21.1 & 19.2 & 20.3 & 22.4 \\
      , 0.01 & $1.14\times10^{41}$ & 25.3 & 20.1 & 18.9 & 19.3 & 19.9 & 18.3 & 18.8 & 21.2 \\
      , 0.01, 0.001 & $1.80\times10^{40}$ & 37.8 & 27.4 & 24.5 & 22.4 & 22.4 & 19.9 & 22.4 & 22.8 \\
      , 0.003 & $5.65\times10^{40}$ & 32.3 & 25.3 & 22.8 & 21.1 & 21.1 & 18.7 & 20.4 & 21.8 \\
      , 0.01 & $1.54\times10^{41}$ & 27.6 & 22.3 & 20.2 & 19.6 & 19.6 & 17.6 & 18.2 & 20.7 \\
      0.45, 0.001, 0.001 & $3.18\times10^{39}$ & 25.5 & 23.3 & 22.2 & 22.9 & 23.5 & 22.4 & 23.5 & 24.1 \\
      , 0.003 & $8.60\times10^{39}$ & 23.2 & 22.1 & 21.5 & 21.9 & 22.5 & 21.5 & 22.5 & 23.3 \\
      , 0.01 & $2.86\times10^{40}$ & 21.5 & 20.7 & 20.3 & 20.5 & 21.3 & 20.6 & 21.2 & 22.5 \\
      , 0.003, 0.001 & $3.40\times10^{39}$ & 29.3 & 23.6 & 22.1 & 22.7 & 23.3 & 22.0 & 23.6 & 24.1 \\
      , 0.003 & $8.90\times10^{39}$ & 24.8 & 22.5 & 21.1 & 21.8 & 22.2 & 20.8 & 22.3 & 23.0 \\
      , 0.01 & $2.63\times10^{40}$ & 22.3 & 21.0 & 20.2 & 20.6 & 21.1 & 19.9 & 20.8 & 22.1 \\
      , 0.01, 0.001 & $3.14\times10^{39}$ & 37.7 & 24.7 & 22.8 & 22.4 & 23.3 & 21.6 & 24.1 & 24.0 \\
      , 0.003 & $9.38\times10^{39}$ & 28.6 & 23.0 & 21.3 & 21.4 & 22.1 & 20.5 & 22.6 & 22.9 \\
      , 0.01 & $2.90\times10^{40}$ & 24.9 & 21.7 & 20.0 & 20.4 & 20.9 & 19.3 & 20.9 & 21.8 \\
      \hline
    \end{tabular}
  }
\end{table}

\nocite{*}
\bibliography{Bibliography}

\end{document}